\documentclass{aa}
\usepackage{natbib}
\bibpunct{(}{)}{;}{a}{}{,}

\usepackage{CJKutf8}
\usepackage{graphicx}
\usepackage{float}

\usepackage{subcaption}
\usepackage{adjustbox}
\usepackage{stfloats}

\usepackage{multirow}
\usepackage{amsmath}
\usepackage{placeins}
\usepackage{amsmath}
\usepackage{caption}
\usepackage{subcaption}
\usepackage{placeins}
\usepackage{float}
\usepackage{multirow}
\usepackage{placeins}
\usepackage{placeins}
\usepackage{color}
\usepackage{appendix}
\titlerunning{Asteroid shape inversion with light curves using deep learning}
\authorrunning{YiJun Tang et al.}

\usepackage[switch]{lineno}

\bibpunct{(}{)}{;}{a}{}{,} 
\usepackage{txfonts}
\usepackage{hyperref}
\hypersetup{
colorlinks=true,
linkcolor=blue,
citecolor=blue,
urlcolor=blue,
}
\nocite{*}
\begin{document}
\captionsetup[figure]{labelfont={bf},labelformat={default},labelsep=period,name={Fig.}}
\captionsetup[table]{labelfont={bf},labelformat={default},labelsep=period,name={Table.}}
\begin{CJK}{UTF8}{gbsn}
\title{Asteroid shape inversion with light curves using deep learning}

   \author{YiJun Tang
          \inst{1,2},
          ChenChen Ying
          \inst{1,2},
          ChengZhe Xia
          \inst{1,2},
          XiaoMing Zhang
          \inst{3}
          \and
          XiaoJun Jiang
          \inst{3} }

   \institute{School of Physics, Zhejiang University of Technology, Hangzhou 310023, China
         \and
         Collaborative Innovation Center for Bio-Med Physics Information Technology of ZJUT, Zhejiang University of Technology, Hangzhou 310023, China
         \and
         CAS Key Laboratory of Optical Astronomy, National Astronomical Observatories, Chinese Academy of Sciences, Beijing 100101, China}
 
  \abstract
   {Asteroid shape inversion using photometric data has been a key area of study in planetary science and astronomical research. Specifically, researchers have focused on developing techniques to reconstruct 3D asteroid shapes from  light curves. This process is crucial for gaining deeper insights into the formation and evolution of asteroids, as well as for planning human space missions. However,  the current methods for asteroid shape inversion require extensive iterative calculations, making the process time-consuming and prone to  becoming stuck in local optima. For missions that aim to make a close approach to an asteroid, a faster and more efficient method is urgently needed.}
   {The goals of this work are to improve the precision, speed, and adaptability to sparse data in asteroid shape inversion and to support autonomous decision-making for shape inversion in space missions. }
   {We directly established a mapping between photometric data and shape distribution through deep neural networks. In addition, we used 3D point clouds to represent asteroid shapes and utilized the deviation between the light curves of non-convex asteroids and their convex hulls to predict the concave areas of non-convex asteroids.}
   {With our approach, we eliminate the need for extensive iterative calculations, achieving millisecond-level inversion speed. We compared the results of different shape models using the Chamfer distance between traditional methods and ours and found that our method performs better, especially when handling special shapes. For the detection of concave areas on the convex hull, the intersection over union (IoU) of our predictions reached 0.89. We further validated this method using observational data from the Lowell Observatory to predict the convex shapes of the asteroids 3337 Miloš and 1289 Kutaïssi, and conducted light curve fitting experiments. The experimental results demonstrated the robustness and adaptability of the method.}
   {We propose a deep learning-based method for asteroid shape inversion using light curve data to reconstruct the convex hull of asteroids and predict concave areas on the convex hull of non-convex asteroids. Our deep learning model efficiently extracts features from input data through convolutional and transformer networks, learning the complex illumination relationships embedded in the light curve data, and enabling precise estimation of the three-dimensional point cloud representing asteroid shapes.}


   \maketitle
%

\section{\textbf{Introduction}} \label{sec:Introduction}

   Asteroids preserve information of the formation and evolution of the Solar System and harbor vast natural resources. In addition, some near-Earth asteroids pose a threat to Earth, making their exploration crucial \citep{2015Icar..257..302H}. The OSIRIS-REx mission targeting the asteroid Bennu \citep{2017SSRv..212..925L} and the Hayabusa2 mission targeting the asteroid Ryugu \citep{2019Sci...364..268W} have already returned their collected samples to Earth, and both spacecraft are continuing their extended missions to new asteroid targets. The Lucy mission, which is targeting the Trojan asteroids \citep{2021PSJ.....2..171L}, was launched and is not designed to return samples. The Tianwen-2 mission\citep{2024LPICo3040.1845Z}, which is planned to launch in 2025, aims to explore the asteroid 1996 FG3 and asteroid 2016 HO3. For all of these missions, ground-based target selection and preliminary shape modeling are essential. However, traditional methods for shape reconstruction have high computational demands, leading to inefficiencies. When faced with the large volumes of data from modern sky surveys[e.g., Large Synoptic Survey Telescope \citep{2019ApJ...873..111I}], traditional iterative methods struggle to process the information effectively. Furthermore, traditional method has inherent uncertainty along the z-axis in the inversion results, limiting their accuracy, and they are prone to becoming stuck in local minima during the optimization process. To address these limitations, the development of deep learning-based algorithms presents a promising alternative. The new approach can significantly enhance inversion speed, improve processing efficiency, and ultimately increase the scientific return of the missions.
   
   Asteroid shape inversion, which is crucial for planetary science research, often relies on photometric data—the easiest and most abundant source for estimating an asteroid's physical characteristics, especially in the absence of high-resolution images. The brightness observed from an asteroid over a time series, known as the light curve, depends on the asteroid’s shape, spin state, and the scattering properties of its surface. This data is essential for improving impact risk assessment, supporting resource exploration, and ensuring successful navigation and landing missions. However, due to the effects of the atmosphere, Earth’s rotation, and the orbital path of asteroids, the effective observation window for ground-based telescopes is limited. While for a small number of larger asteroids, it is easier to obtain high signal-to-noise ratio data and high-resolution images, most asteroids have sparse historical photometric data, thus complicating data processing.  Despite these challenges, photometric data remains invaluable for providing essential information for simulations, modeling, and further understanding of asteroid characteristics\citep{2023A&A...675A..24D, 2020A&A...643A..59D}.

   \citet{1906ApJ....24....1R} proposed that observing the light curve only during opposition geometry is insufficient to determine the shape of an asteroid. Nearly a century later, new paradigms and theories for asteroid light curve inversion were introduced and established led by \citet{1992A&A...259..333K} and \citet{1992A&A...259..318K}. The authors demonstrated that light curve data observed under different geometries of the Sun-asteroid-observatory relationship could be used to estimate the Gaussian curvature of asteroids. Furthermore, they proposed preliminary numerical inversion methods for retrieving convex body models, which were successfully applied by \citet{1992A&A...266..385B} in the 3D shape inversion of asteroid(951) Gaspra. Ten years later, \citet{2001Icar..153...24K} and \citet{2001Icar..153...37K} established a mature and robust theory and methodology for asteroid shape inversion using convex shape models (which we refer to as KTM in this paper). The quality of the inversion method heavily relies on the quality of light curves—they should be fairly dense, cover the entire spin cycle, and be observed under different solar-asteroid observer geometries.
   
   To exploit sparse light curve data, \citet{2012P&SS...73...80C} studied a simpler shape representation for asteroids during the inversion process with fewer parameters. Their studies on several asteroids indicated that the spin poles and periods recovered using triaxial ellipsoids were essentially consistent with those from the KTM solution with complex shapes. By integrating the Lommel-Seeliger surface scattering model, \citet{2015A&A...584A..23M} further advanced the ellipsoidal asteroid model. Following the same line of work, \citet{2015P&SS..118..227M} employed a Markov Chain Monte Carlo (MCMC) analysis to describe the probability density function of the neighborhood region of the best-fitting parameters. \citet{2010A&A...513A..46D} established a three-dimensional model database website containing 16091 models of asteroids: Database of Asteroid Models from Inversion Techniques (DAMIT). The database also offers the convex model inversion program by \citet{2001Icar..153...24K}, which has now become one of the most popular inversion tools. Many of the models in the database are convex inversion models derived from sparse photometric data\citep{2010A&A...513A..46D, 2020A&A...643A..59D, 2024A&A...687A.277C}.

   Most of the shape models generated through light curve inversion are convex, primarily because the addition of convexity constraints during the inversion process can stabilize the solution. As noted by \citet{2001Icar..153...24K}, although asteroids typically possess concavities, their light curves can be approximated by those generated from convex shape models. Convex models serve as alternative solutions to non-convex models lacking mathematical uniqueness and stability. They are only necessary when dealing with high-quality light curves at high solar phase angles or disk-resolved data. \citet{2003A&A...404..709D} noted that at high solar phase angles, the shadowing effects caused by non-convex models become more pronounced on the light curve. \citet{2018MNRAS.473.5050B} developed the SAGE method based on genetic algorithms, and it relaxes the typical convex assumption during the light curve inversion process. The SAGE method iteratively generates random shape and spin axis mutations, evaluating them at each step until a stable solution is found. This method is particularly useful when working with high-quality light curves at high solar phase angles or disk-resolved data, where non-convexity becomes more critical to accurately modeling asteroid shapes.

   \citet{kaasalainen2011multimodal} proposed a weighting strategy to combine various types of observational data for asteroid shape, scattering parameters, rotation parameters, and albedo distribution inversion. \citet{2012A&A...543A..97K} achieved multiple complementary data inversions. Subsequently,  \citet{2012P&SS...66..200C} introduced Knitted Occultation, Adaptive-optics, and Lightcurve Analysis(KOALA), aiming to integrate light curve inversion with adaptive optics and stellar occultation. These methods were further developed by \citet{2015A&A...576A...8V}, who proposed All-Data Asteroid Modeling(ADAM) algorithm to combine various types of data. These methods involve not only optical photometric data but also infrared data, radar data, occultation data, interferometric data, adaptive optics data, and high-resolution imaging data. \citet{2021A&A...649A..98M} devised a method to extract reference absolute magnitudes and phase curves from Gaia data, enabling comparative analyses across hundreds of asteroids. Furthermore, \citet{2022FrASS...9.1125M} introduced error models for various light curve categories and investigated the behavior of phase angles.

   Asteroid shape inversion based on light curve data has long been treated as a type of non-convex optimization problem\citep{danilova2022recent}, characterized by challenges such as multiple local minima arising from the mathematical complexity of the inversion process. Traditionally, this problem is tackled using deterministic initialization and numerical iterative minimization methods. Aiming to achieve a global solution, \citet{2022MNRAS.513..311C} proposed a global optimization framework by embedding the spin pole and area vector determination module. While theoretically more effective, this approach nonetheless remains computationally intensive and is limited to convex body models, thus making it unsuitable for more complex non-convex shapes.
   
    In recent years, there has been a trend towards treating shape estimation of space objects as a classification problem and employing deep learning methods to address it. \citet{2020JAnSc..67.1063L} trained a convolutional neural network (CNN) solely based on light curve data to classify space objects by type and included categories such as rocket bodies, payloads, and debris. Their method achieved an accuracy of 75$\%$ on real light curve test sets, demonstrating the feasibility of using neural networks to process light curve data. \citet{2021AcAau.181..301A} demonstrated that transferring knowledge learned on simulated light curve data improved the performance of deep networks in shape classification tasks using real light curve data. However, the application of deep neural networks in asteroid light curve-related tasks has stalled at this point, with no further successful application of neural networks to three-dimensional shape inversion. 

   Advancements in 3D sensor technology have significantly propelled research within the 3D computer vision domain. Among the various 3D data formats, point clouds are particularly popular due to their ability to efficiently store detailed 3D shape information while consuming less memory than other formats such as voxel grids or 3D meshes. Nonetheless, the point cloud data captured by current 3D sensors often suffers from incompleteness and inaccuracies, stemming from challenges such as self-occlusion, light reflection, and the inherent limitations in sensor resolution. As a result, the task of reconstructing complete point clouds from partial and sparse datasets has become increasingly crucial and is now considered essential in this field. Researchers have explored various approaches within deep learning to address this challenge. Initial attempts in point cloud completion sought to adapt established methods from 2D completion tasks to 3D point clouds\citep{dai2017shape,han2017high,sharma2016vconv,stutz2018learning,nguyen2016field}, primarily through voxelization and 3D convolutions.

  With the success of PointNet \citep{qi2017pointnet} and PointNet++ \citep{qi2017pointnet++}, directly processing 3D coordinates has become the mainstream technique for point cloud-based 3D analysis. This technology has been further applied to many pioneering works in point cloud completion, where encoder-decoder architectures are designed to generate complete point clouds \citep{achlioptas2018learning, tchapmi2019topnet, groueix2018papier, mandikal2019dense}. These algorithms have proven to be particularly well-suited for tasks in autonomous driving, especially with data generated by LiDAR sensors \citep{yu2021pointr, zhou2022seedformer, hao2023improved, hartmann2024pointnet}, as they excel at extracting both global and local features from sparse point cloud data. Additionally, earlier approaches that focused on voxelization and 3D convolutions for point cloud completion \citep{yuan2018pcn, huang2020pf}, have laid the groundwork for the development of more sophisticated methods.

   Points clouds are used to represent the surface shape of three-dimensional objects. A point cloud consists of coordinates (x, y, z), and the collection of these points forms the semantic features of a three-dimensional object. The approach of representing 3D shapes with point clouds has inspired us. In the task of three-dimensional shape inversion based on light curve data, the brightness recorded at each sampling point, along with the corresponding coordinates of the Sun and the observatory at each sampling point, forms an independent vector. The collection of many such vectors embeds information about the asteroid's 3D shape features. We aim to leverage deep learning techniques to learn the shape information embedded in the brightness data while utilizing large datasets for training and learning. Compared to iterative optimization methods, regression-based prediction approaches can significantly improve the efficiency of shape inversion.

\section{\textbf{Preliminaries}} \label{sec:floats}
\subsection{Asteroid light model} \label{subsec:Asteroid light model}

   We considered an asteroid in principal-axis rotation about its axis of maximum inertia and denoted the rotation period by T, the pole orientation in ecliptic longitude and latitude by ($\lambda, \beta)^r$ (J2000.0; here, $r$ stands for transpose), and the rotational phase at a given epoch $t_0$ by $\phi_0$. As to the asteroid shape, we incorporated either triaxial ellipsoidal shapes or general convex shapes.

The observed brightness $I$ of the asteroid at any given time $t$ can be expressed as:

   \begin{equation}
   I(t) = \frac{F_{in}}{\left( v \right._{obj}^{i})^{2}} \sum_{\substack{i=1 \\ s_i > 0}}^{nFac} S_i(\mu, \mu_0, \alpha) A_i,
   \end{equation}

   Where $F_{in}$ is the incident flux density, $v_{obj}^{i}$ represents the distance from the asteroid to the observatory, and $nFac$ represents the number of surface facets into which the asteroid's surface is divided. The term $S\left( {\mu,\mu_{0},\alpha} \right)$ represents the surface scattering coefficient, and it is a function of $t$, where $\mu$, $\mu_{0}$, and $\alpha$ correspond to the cosine of the angle between the observer's direction and the asteroid's surface normal, the cosine of the angle between the light source direction and the asteroid's surface normal, and the phase angle, respectively. Here, $s_i > 0$  indicates that the $i-th$ facet is simultaneously illuminated by the Sun and visible from the observatory at time $t$, and $A_i$ represents the area of the $i-th$ facet.
  
    Scattering models serve to describe the statistical characteristics of how light is scattered, reflected, and absorbed by the microscopic structures on asteroid surfaces. Numerous scattering models exist today, and each is dependent on variables such as incidence angle, emission angle, phase angle, surface roughness, porosity, particle size, and other physical parameters\citep{2021arXiv210601363R}. Among the commonly used scattering models for asteroids are the LS-L (Lommel-Seeliger-Lambert) model introduced by Kaasalainen and the model developed by Hapke\citep{1963JGR....68.4571H,1966AJ.....71..333H}.
    
    The Hapke model encompasses a range of physical characteristics, including the incidence angle, emission angle, average single-scattering albedo, scattering and extinction coefficients of the medium, detector sensitivity area, reflectance, the number of reflective particles per unit volume, average particle cross-sectional area, average path length of light attenuation within the medium, and surface roughness. This model also accounts for phenomena such as multiple scattering, mutual shadowing, opposition effects across different wavelengths, and anisotropic scattering. Due to its complexity, the Hapke model has been applied to the inversion of only a limited number of asteroids\citep{2021AJ....161..112L,2010M&PS...45.1804S,2002Icar..159..192D,2014Icar..233...48P,2015Icar..257..239M,1998Icar..135..451H}, with the light curve fitting residuals showing limited sensitivity to the model's parameters. Moreover, extracting all surface scattering characteristics from optical photometric data remains a significant challenge\citep{2015aste.book..129L}.
    
    \citet{2001Icar..153...24K} pointed out that the scattering model used for inversion needs to be sufficiently simple, as having fewer model parameters can enhance the stability of the inversion. The same applies to neural networks—fewer model parameters are more conducive to the network learning the shape features from light curve data. So,we adopted the linear combination of Lommel-Seeliger (LS) and Lambert (L) models \citep{2001Icar..153...37K} as the scattering law of the object, formulated as:
   \begin{equation}
   \begin{split}
   S\left( {\mu,\mu_{0},\alpha} \right) &= f(\alpha)\left\lbrack {S_{LS}\left( {\mu,\mu_{0}} \right) + {cS}_{L}\left( {\mu,\mu_{0}} \right)} \right\rbrack \\
   &= f(\alpha)\left\lbrack \frac{\mu\mu_{0}}{\mu + \mu_{0}} + c\mu\mu_{0} \right\rbrack ,
   \end{split}
   \end{equation}
   
   where the $f(\alpha)$ is the phase function
   \begin{equation}
   \label{eq:3}
   f(\alpha) = aexp\left( {- \frac{\alpha}{d}} \right) + k\alpha + 1.
   \end{equation}

   and the terms $a$ and $d$ are the amplitude and scale length of the opposition effect, $k$ is the overall slope of the phase curve, and $c$ is a weight parameter.

\subsection{Shape representation} \label{subsec:Shape representation}

   In the field of computer vision, three-dimensional objects can be represented in various ways, but primarily it is done with meshes\citep{catmull1974subdivision}, voxels\citep{lorensen1998marching}, and point clouds. The mesh encodes the geometry of three-dimensional objects based on the combination of vertices, edges, and faces, often using polygonal faces to represent the surface of three-dimensional objects. Triangular grids are one of the simplest types of grids and are commonly used for the convenient and rapid generation of three-dimensional surface representations, particularly in unstructured grids, where flexibility and efficiency are essential.

   Voxels are the corresponding representation of pixels in a three-dimensional space (likewise, pixes are analogous to the representation of voxels in two-dimensional images). Voxelization is the process of converting a continuous geometric object into a set of discrete voxels that closely approximate the object. Voxels can be viewed as cubes representing uniformly spaced samples on a three-dimensional grid, and they are commonly used in volume rendering, medical image processing, fluid dynamics simulation, and other fields, especially for dense three-dimensional data.

   Point clouds are collections of data points in three-dimensional space, which can be used to describe the geometric shape of individual objects or entire scenes. Each point in a point cloud is defined by X, Y, and Z coordinates, representing the physical position of the point in three-dimensional space. Point clouds are commonly used in geographic information systems (GIS), three-dimensional modeling, virtual reality, autonomous driving, and other fields, especially for non-uniform or discrete three-dimensional data. Due to the lack of explicit topological information, point clouds offer greater flexibility in usage than traditional methods such as meshes or voxels.

   We represent the three-dimensional shape of the asteroid as a collection of point clouds, denoted as P, which is a discrete representation of its underlying continuous surface. For traditional point cloud collections, each point represents only a three-dimensional coordinate, and its capability to express the three-dimensional shape depends on the resolution of the point cloud.

\subsection{Inverse problem}

   The inversion of asteroid shapes from light curves fundamentally relies on the geometric relationships between the observer, the Sun, and the asteroid's surface. At a certain observation moment $t_i$, the brightness $I_i$ of the asteroid is observed, and the direction vector of the Sun in the asteroid's body coordinate system $v_{sun}^{i}=(x_{sun}^{i},y_{sun}^{i},z_{sun}^{i})$ and the direction vector of the observatory $v_{obj}^{i}=(x_{obj}^{i},y_{obj}^{i},z_{obj}^{i})$ are recorded. Assuming the direction vector in the ecliptic coordinate system is $v_{ecl}^{i}$, the transformation formula of the body coordinate $v_{ast}^{i}$ with the asteroid as the center is then
    \begin{equation}
  v_{ast}^{i} =  R_{z}(\theta)R_{y}(90^{\circ} - \beta)R_{z}(\lambda)v_{ecl}^{i}.
     \end{equation}
   Here, $ R_{z}(\theta)$, $ R_{y}(90^{\circ} - \beta)$, and $R_{z}(\lambda)$ are  rotational matrices, where $\theta$  denotes the rotational phase angle
   
   \begin{equation}
  R_{z}(\theta) = \left( \begin{matrix}
    {\cos(\theta)} \\
    {- {\sin(\theta)}} \\
    0
    \end{matrix} \right.~~~~~\begin{matrix}
    {sin(\theta)} \\
    {cos(\theta)} \\
    0
    \end{matrix}~~~~~\left. \begin{matrix}
    0 \\
    0 \\
    1
    \end{matrix} \right)
    \end{equation}

    \begin{equation}
    R_{y}(90^{\circ} - \beta) = \left( {\begin{matrix}
    {\sin(\beta)} \\
    0 \\
    {\cos(\beta)}
    \end{matrix}~~~~~} \right.\begin{matrix}
    0 \\
    1 \\
    0
    \end{matrix}\left. {~~~~~\begin{matrix}
    {- {\cos(\beta)}} \\
    0 \\
    {\sin(\beta)}
    \end{matrix}} \right)
    \end{equation}
    
   \begin{equation}
   R_{z}(\lambda) = \left( \begin{matrix}
    {\cos(\lambda)} \\
    {- {\sin(\lambda)}} \\
    0
    \end{matrix} \right.~~~~~\begin{matrix}
    {sin(\lambda)} \\
    {cos(\lambda)} \\
    0
    \end{matrix}~~~~~\left. \begin{matrix}
    0 \\
    0 \\
    1
    \end{matrix} \right)
    \end{equation}

    \begin{equation}
    \theta = \theta_{0}+\frac{2\pi}{T}(t-t_{0}).
     \end{equation}
    
   Here,  $\lambda$  represents the ecliptic longitude, $\beta$ is the ecliptic latitude of the asteroid's rotation axis, $t_{0}$ denotes the initial time, the parameter $\theta_{0}$ represents the asteroid's rotational phase at that moment, and $T$ indicates the period of the asteroid's rotation.
   
    Recording the brightness at time \( t_i \) along with the normalized direction vectors of the Sun and the observatory in the asteroid's body coordinate system yields a vector representation. We let the observation vector at time \( t_i \) be denoted as \( q_i = (I_i, x_{\text{sun}}^i, y_{\text{sun}}^i, z_{\text{sun}}^i, x_{\text{obj}}^i, y_{\text{obj}}^i, z_{\text{obj}}^i) \), where \( (x_{\text{sun}}^i, y_{\text{sun}}^i, z_{\text{sun}}^i) \) and \( (x_{\text{obj}}^i, y_{\text{obj}}^i, z_{\text{obj}}^i) \) are the normalized direction vectors of the Sun and the observatory, respectively, in the asteroid's body coordinate system. The collection of observation vectors at various historical moments forms a set 
    $
    Q = \left\{ {q_{i} \in \mathbb{R}^{7}} \right\}_{i = 1}^{M}
    $. To ensure that our inversion process only has shape as a variable, we processed $I_i$ by assuming the phase function is known. Before inputting each point into the network, we divided $I_i$ by the phase function $f(\alpha)$ as given in Equation~\eqref{eq:3}. To further normalize $I_i$, we divided each $I_i$  after the phase function correction by the average value of all $I_i$ in the set $Q$. We refer to the observation vector set $Q$ after this preprocessing as $Q_{preprocess}$.
    
    At this point, we needed to establish a mapping relationship $\phi$:
    \begin{equation}\label{eq:6}
    \left. \phi : Q_{preprocess} \longrightarrow P \right. \,,
    \end{equation}

\begin{figure*}[ht]
    \centering
    \includegraphics[width=1.0\textwidth]{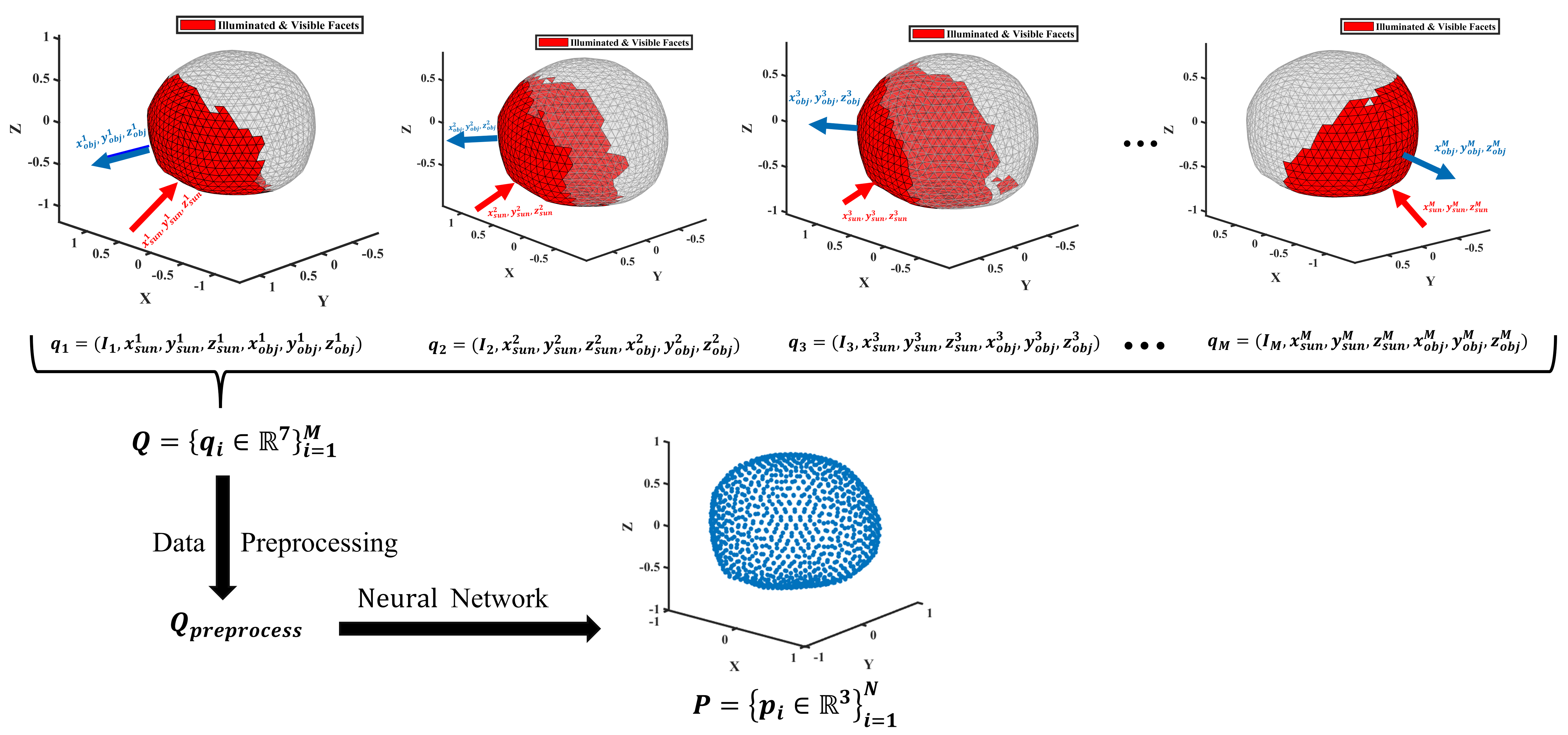}
    \captionsetup{justification=justified, singlelinecheck=false}
    \caption{Illustration of our proposed inversion method. The top four panels show the schematic diagrams of different visible and illuminated facets of the asteroid under varying Sun-asteroid-Earth positional relationships in the asteroid's body coordinate system. The blue point cloud represents the asteroid point cloud obtained from the inversion process.}
    \label{fig:approach}
\end{figure*}

    Where $P$ represents the point cloud set characterizing the three-dimensional shape of the asteroid. The construction of $\phi$ was implemented using a neural network. An illustration of our inversion method is presented in Fig.\ref{fig:approach}.
    
  At the same time, we noticed that the area illuminated and observable on the asteroid during two independent observations can  overlap, and this information is crucial for the inversion process. The occurrence of such area overlap is related to the shape of the asteroid and the direction of illumination. Such situations are likely to occur when the observation vectors $v_{obj}^{i}$ and ${v}_{obj}^{j}$ in the asteroid's body coordinate system are very close. This can happen both in observations from different historical periods and in continuous observations of the asteroid during a single night. By normalizing ${v}_{obj}^{i}$, we obtained ${v'}_{obj}^{~i}$. Hence, we could define a domain $\mathcal{D}_{i} \subset \mathbb{R}^{3}$ centered at the point ${v'}_{obj}^{~i}$ with a fixed radius $r$:
    \begin{equation}
    \mathcal{D}_{i} = \left\{ {v^{'}}_{obj}^{~i},\left| {{v'}_{obj}^{~i} - {v'}_{obj}^{~j}} \right| < r \right\}.
    \end{equation}
    
    Here,  the $\left| {v'}_{obj}^{~i} - {v'}_{obj}^{~j} \right| $ represents the Euclidean distance between the ${v'}_{obj}^{~i}$  and ${v'}_{obj}^{~j} $. Due to the unknown shape of the asteroid, it is not possible to determine an accurate value of $r$ in order to judge whether there is an overlap in the observed areas. Therefore, in our experiment, $r$ was set as an empirical value, which we set as 0.1. In the subsequent feature extraction process, appropriate feature fusion was performed on $\mathcal{D}_{i}$.
    
\section{\textbf{Proposed method}}\label{sec:Proposed Method}
    \subsection{Convex  inversion}

    The goal of our proposed algorithm is to establish the mapping relationship as indicated in Equation~\eqref{eq:6}. However, we must state first that the method is based on the known parameters of rotation axis direction, rotation period, and phase function.
    
    We trained the neural network on a vast amount of simulated data, attempting to establish this mapping relationship. We provide a comprehensive overview of our methodology later in this work.
    
    We made improvements to the PoinTR{\citep{yu2021pointr}} algorithm framework. The network divides the prediction process of asteroid point clouds into two stages. The first stage is coarse point cloud prediction, which predicts the basic outline of the three-dimensional shape, and the second stage further refines the surface of the coarse point cloud.

\begin{figure*}[ht]
    \centering
    \includegraphics[width=1.0\textwidth]{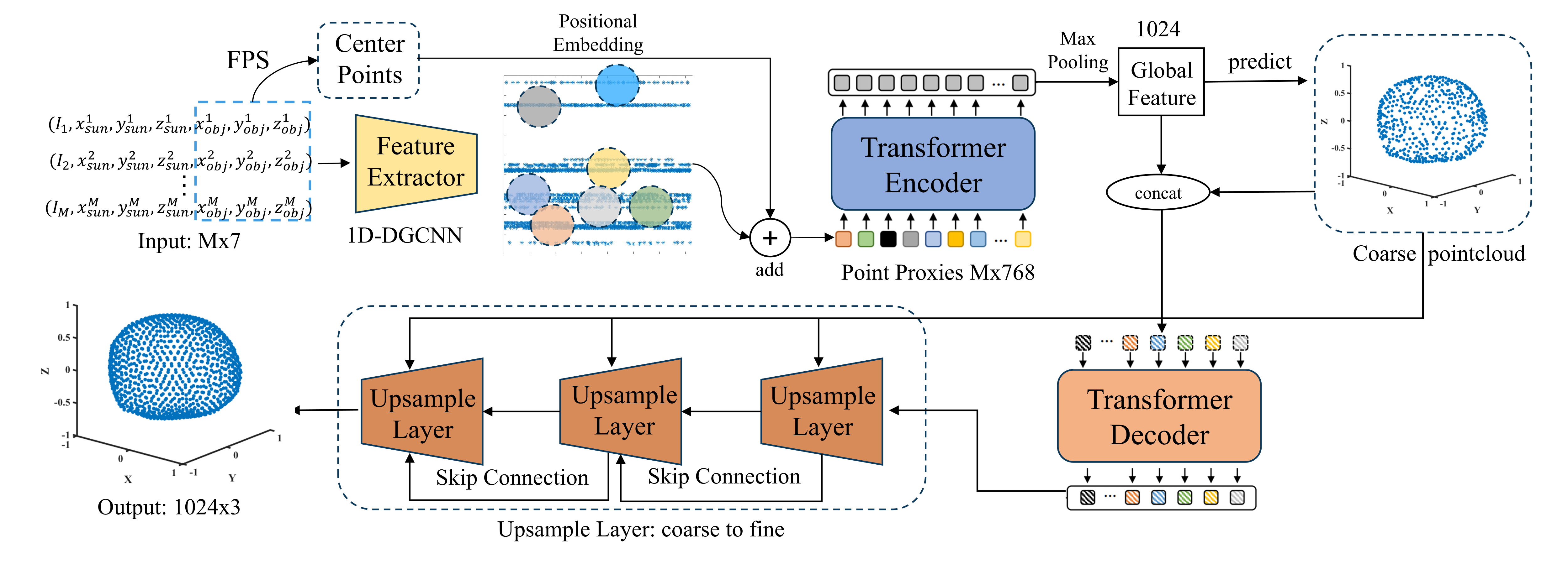}
    \captionsetup{justification=justified, singlelinecheck=false}
    \caption{Network architecture. First, FPS is used to downsample the input light curve sampling points in order to determine the center points. Then, the nearest neighbor set corresponding to each center point is obtained with the KNN algorithm, and local features are extracted using a simplified 1D-DGCNN. Afterwards, positional embeddings are added to enhance spatial awareness. The enhanced features are encoded by a geometry-aware transformer module followed by global feature aggregation through max pooling. Finally, the coarse point cloud is refined into the output using a transformer decoder that integrates three up-sampling layers and skip connections.}
    \label{fig:f1}
\end{figure*}

    Following the style of PoinTR\citep{yu2021pointr}, We used a simplified 1D-DGCNN\citep{wang2019dynamic} network for feature extraction of input vectors. We employed farthest point sampling(FPS) and the k-nearest neighbors (KNN) algorithm to extract local features and downsample them. Furthermore a geometry-aware transformer module \citep{yu2021pointr} was utilized  for feature encoding. In this module, the multi-head self-attention and KNN query modules work in parallel, capturing both global information and local information of light curves. The output features pass through a fully connected layer and a max-pooling layer to obtain global features, and based on these features, the coarse point cloud of the three-dimensional shape is predicted. Subsequently, we input the coarse point cloud along with the encoded features into the decoding stage. The structural features of the coarse point cloud were decoded along with the features input into the network. Finally, we performed three layers of point cloud upsampling based on the UpSample transformer module proposed by \citet{zhou2022seedformer}. In this module, the coarse point cloud is processed along with the decoder outputs through two branches: one that preserves the original features and another that refines them using attention weights computed from local neighborhoods. These refined features were then used to upsample the point cloud while retaining the geometric structure and completing the shape based on the original light curve features. The refined point cloud was obtained as the final output. The network architecture is shown in Fig.\ref{fig:f1}. The specific structure of the network can be found in the supporting documents.

    In the experiment, the Chamfer distance (CD) was employed as the loss function of the network to measure the distance between two unordered point sets, defined as follows:
    \begin{equation}
    \begin{split}
    d_{CD}\left( {P,Q} \right) = &\frac{1}{S_{1}} \sum\limits_{x \in S_{1}} \min_{y \in S_{2}} \left\| {x - y} \right\|_{2}^{2} \\
    + &\frac{1}{S_{2}} \sum\limits_{y \in S_{2}} \min_{x \in S_{1}} \left\| {x - y} \right\|_{2}^{2}. \label{eq:10}
    \end{split} 
    \end{equation}

    Where $S_1$ and $S_2$ represent two sets of point clouds, and $x$, $y$ respectively denote the three-dimensional vectors of coordinates for the two sets of point clouds.

    To ensure that both the generation of coarse point clouds and the progressive refinement approach closely approximate the real shape, the output of each point cloud at every stage was constrained using the CD loss. The coarse point cloud underwent a total of three upsampling steps before producing the final result. Therefore, the total loss function $L_{pc}$ is
    \begin{equation}
    L_{pc} = L_{coarse} + L_{c2f} + L_{fine}.
    \end{equation}

    Here, $L_{coarse}$represents the chamfer loss of the coarse point cloud, $L_{c2f}$ represents the chamfer loss of the output point cloud after the first layer of upsampling, and $L_{fine}$ represents the chamfer loss of the final refined point cloud output.
    
    \subsection{Determination of concave areas}
    
    \citet{1992A&A...259..333K} pointed out that although  asteroids typically have concavities, their light curves are approximately similar to those generated by convex shapes. Non-convex models serve as an alternative to convex models; however, their solutions lack mathematical uniqueness and stability unless supplemented by additional sources of shape information. By minimizing the photometric error, the shape we invert is very close to the convex hull of the non-convex asteroid. \citet{2015MNRAS.453.2232D} utilized this feature to propose a method for predicting concave areas by analyzing the local normal distribution of the model. However, their method does not use the photometric information of the preliminary prediction results further and can only detect large-scale concavities. Our aim is to develop a new method for detecting non-convex regions utilizing the photometric information from both the model and the convex hull, and analyzing their differences (even when the differences are very small under low solar phase angles or when the non-convexity is minimal), along with the corresponding positions of the Sun and the observation station in order to determine the non-convex regions.
    
    Thus, we  utilized the deviation between the light curve of the  convex hull and the actual light curve, along with the illumination position information, to determine the concave areas and the degree of concavity. At this point,  the input observation vector becomes
    $q_{i}^{'} = \left( I_{c}^{i},I_{nc}^{i},I_{diff}^{i},x_{sun}^{i},y_{sun}^{i},z_{sun}^{i},x_{obj}^{i},y_{obj}^{i},z_{obj}^{i} \right)$, forming a similar set $Q^{'} = \left\{ {q_{i}^{'} \in \mathbb{R}^{9}} \right\}_{i = 1}^{M}$,where $I_{c}^{i}$ represents the relative brightness of the convex hull, $I_{nc}^{i}$ represents the relative brightness of the  non-convex model, and $I_{diff}^{i}$ represents the difference between the brightness generated by the convex hull simulation and the brightness of the non-convex model.
    
\begin{figure}[!ht]
        \centering
        \includegraphics[width=0.5\textwidth]{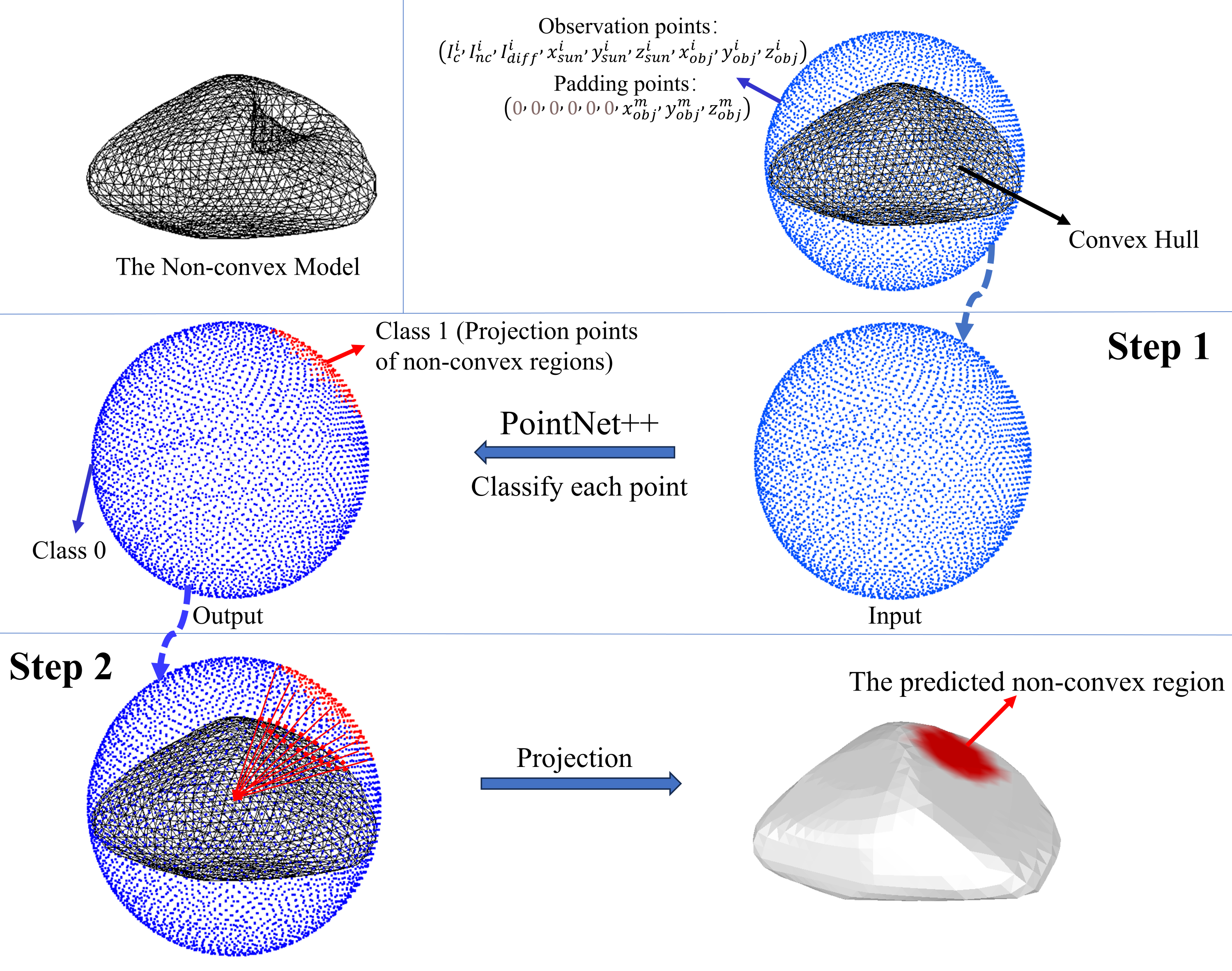}
        \captionsetup{justification=justified, singlelinecheck=false}
        \caption{Method of predicting non-convex regions.}
        \label{fig:n1}
    \end{figure}

    As our aim is for the information output by the neural network to express the non-convex regions on the convex hull, we therefore performed some pre-processing on $Q^{'}$ before feeding the data into the network. We employed the Fibonacci sphere sampling method \citep{brent2013algorithms} to complete the normalized observation vectors ($x_{obj}^{i},y_{obj}^{i},z_{obj}^{i}$). This ensures that the normalized observation points always form a spherical shape in space, with the photometric and illumination information for the supplemented points entirely replaced by zeros, such as (0,0,0,0,0,0,$x_{obj}^{i}, y_{obj}^{i}, z_{obj}^{i}$). These supplementary vectors are added to $Q^{'}$ to form a new set, $Q^{''}$, which is then fed into the classification network based on PointNet++ \citep{qi2017pointnet++}. The network classifies each vector and identifies which points represent the projection of the non-convex regions, as shown in Fig.\ref{fig:n1}. This projection method uniquely represents the concave areas within a 3D model, and our experiments demonstrated the reliability of this approach.

    \section{\textbf{Results and discussion}}
        \subsection{Simulated light curve experiments}
\begin{figure}[]
        \centering
        \includegraphics[width=0.5\textwidth]{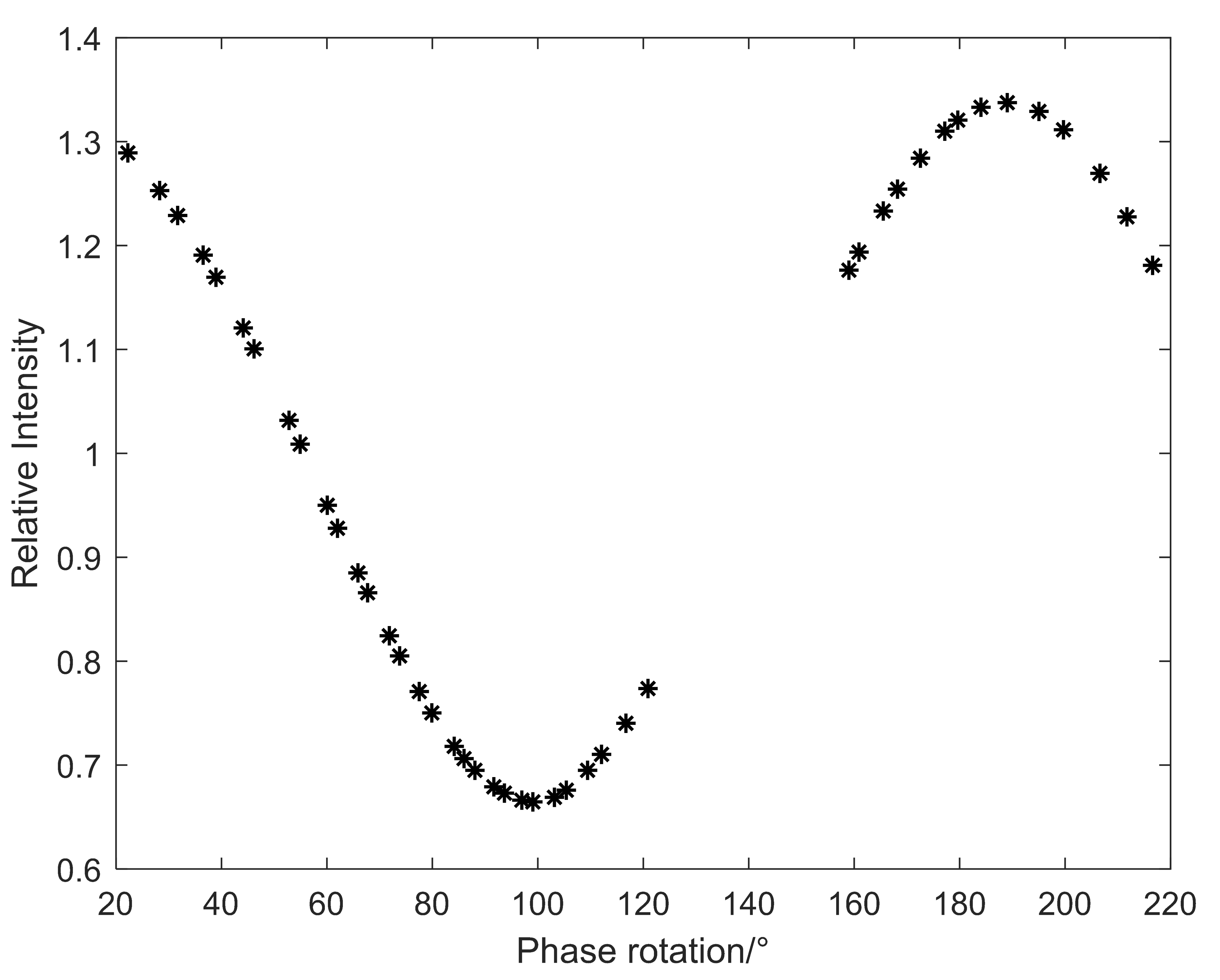}
        \captionsetup{justification=justified, singlelinecheck=false}
        \caption{Simulated light curve for convex.}
        \label{fig:f3}
    \end{figure}

    Due to the influence of an asteroid's orbital motion and Earth's orbital motion, each asteroid can only be observed during certain limited seasons. To ensure that the simulated observation conditions are realistic, we used the geometric position relationship between the Sun, the asteroid, and the observatory from actual data provided by the DAMIT website. Each asteroid's historical observation positions are different, and by combining the 3D models of multiple asteroids with various actual observation positions, a vast number of light curves can be generated. We present one of them in the Fig.\ref{fig:f3}. Essentially, this generation method assigns different orbital paths to each asteroid model. Since the simulation uses actual observation position data, the generated light curves better reflect real observation conditions, further enhancing the realism of the simulation method.

    We used 3,459 3D asteroid models to simulate light curves. For each 3D asteroid model, we selected the orbital information of 35 asteroids (the Sun-asteroid-Earth positional relationships) to simulate and generate light curve samples. Each light curve sample corresponds to a set of $Q_{preprocess}$, as mentioned in Section 2.3. The number of light curves within a sample depends on the number of historical light curves of the selected asteroid used to determine the Sun-asteroid-Earth positional relationships during the simulation, and it ranges from two to eight light curves per sample. In total, we generated 121,065 light curve samples, which were divided into training and testing sets at a ratio of approximately 10:1.

\begin{figure}[!ht]
        \centering
        \includegraphics[width=0.5\textwidth]{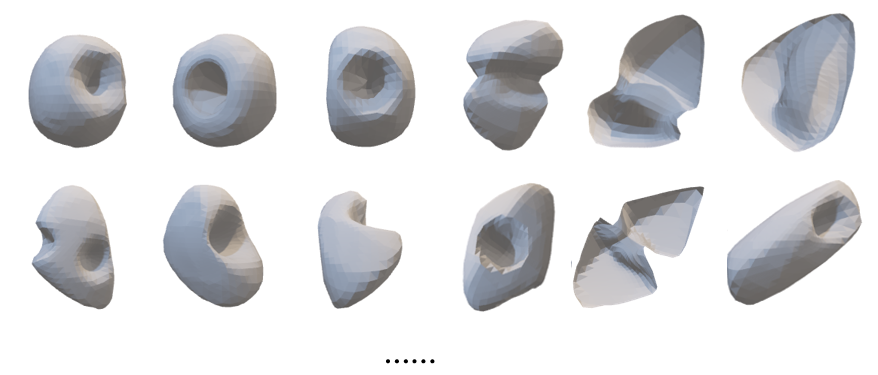}
        \captionsetup{justification=justified, singlelinecheck=false}
        \caption{Part of non-convex 3D model datasets.}
        \label{fig:n2}
    \end{figure}

    For the non-convex shape, we added a conditional check for facet occlusion in the light curve simulation. Using a triangle intersection detection algorithm\citep{moller2005fast}, we traversed each facet to determine whether it was occluded by other facets in the direction of illumination and observation. We used Blender\citep{flavell2011beginning} software to make 1360 non-convex models with different depression degrees and different depression positions; part of the non-convex 3D models are shown in  Fig.\ref{fig:n2} . For each non-convex model and its convex hull model, we selected the orbital information of 20 asteroids and used the recorded Sun-asteroid-Earth positional relationships to simulate and generate light curve samples. We simulated 27,200 light curve samples for the non-convex model and its convex hull model under actual observation conditions. Because 3D models of non-convex asteroids are difficult to obtain, our dataset contains relatively fewer non-convex models. Since neural network training requires large amounts of data, we allocated more data to the training set, dividing the training and testing sets at a ratio of approximately 14:1. Specific dataset information is shown in Table.\ref{tab:simulated_data_distribution}.

The algorithms discussed in this paper were implemented in Python and trained on a laboratory server running Ubuntu 20.04.1. The network architecture was built using the PyTorch library, with deep learning acceleration provided by the NVIDIA CUDA library. The server is equipped with 8 RTXA5000 GPUs, each with 24GB of video memory, 512GB of RAM, and two 10-core, 16-thread CPUs operating at 2.90GHz. The models were trained and tested across these GPUs for a total of 400 epochs, using the AdamW optimizer with learning rate adjustments managed by the LambdaLR scheduler. The batch size was set to 64, and the initial learning rate was 0.001.
    
\begin{table*}[!htbp]
    \centering
    \captionsetup{justification=justified, singlelinecheck=false}
    \caption{Dataset distributions.}
    \small  
    \begin{tabular*}{\textwidth}{@{\extracolsep{\fill}}lcccc}
        \hline
        & \multicolumn{2}{c}{Training Set} & \multicolumn{2}{c}{Test Set} \\
        & Number of Models & Number of light curve samples & Number of Models & Number of light curve samples \\
        \hline
        Convex Inversion  & 3113 & 108955 & 346 & 12110 \\
        Determination Of Concave Areas & 1270 & 25400 & 90 & 1800\\       
        \hline
    \end{tabular*}
    \label{tab:simulated_data_distribution}
\end{table*}

   \subsection{Convex inversion results}
\begin{figure}[htbp]
        \centering
        \includegraphics[width=0.5\textwidth]{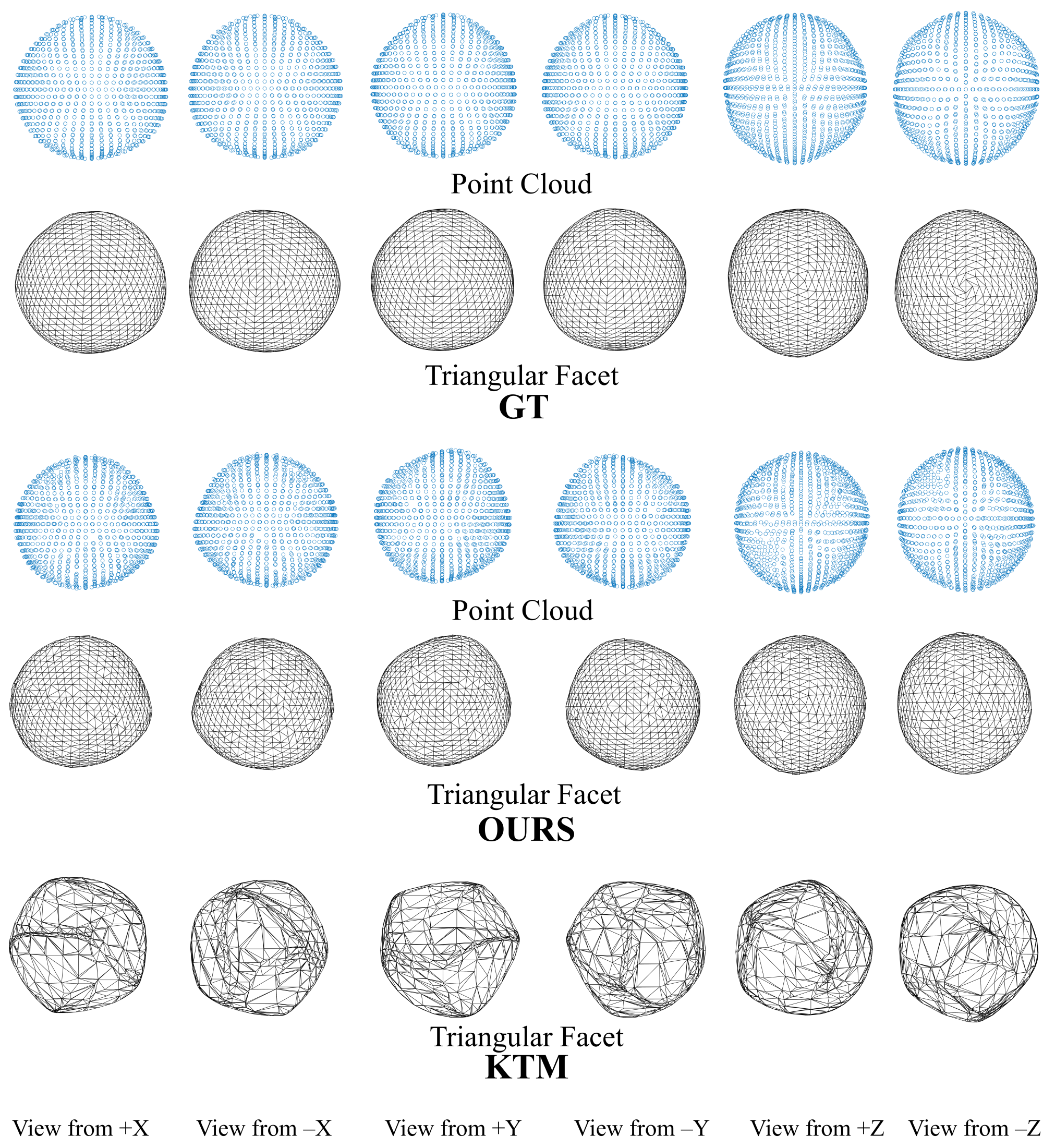}
        \captionsetup{justification=justified, singlelinecheck=false}
        \caption{Inversion results of spherical model. The models are divided into three groups. The first group represents the ground truth of the 3D model, displayed using both point cloud and facet representations. The second group shows the 3D model reconstructed using our method, where the point cloud is converted into triangular facets for easier comparison with the KTM method. The third group represents the 3D model reconstructed using the KTM method. All models are aligned with the asteroid-centric coordinate system for consistency and clarity in comparison.
        \label{fig:qiu4}
        }
    \end{figure}

\begin{figure}[htbp]
        \centering
        \includegraphics[width=0.5\textwidth]{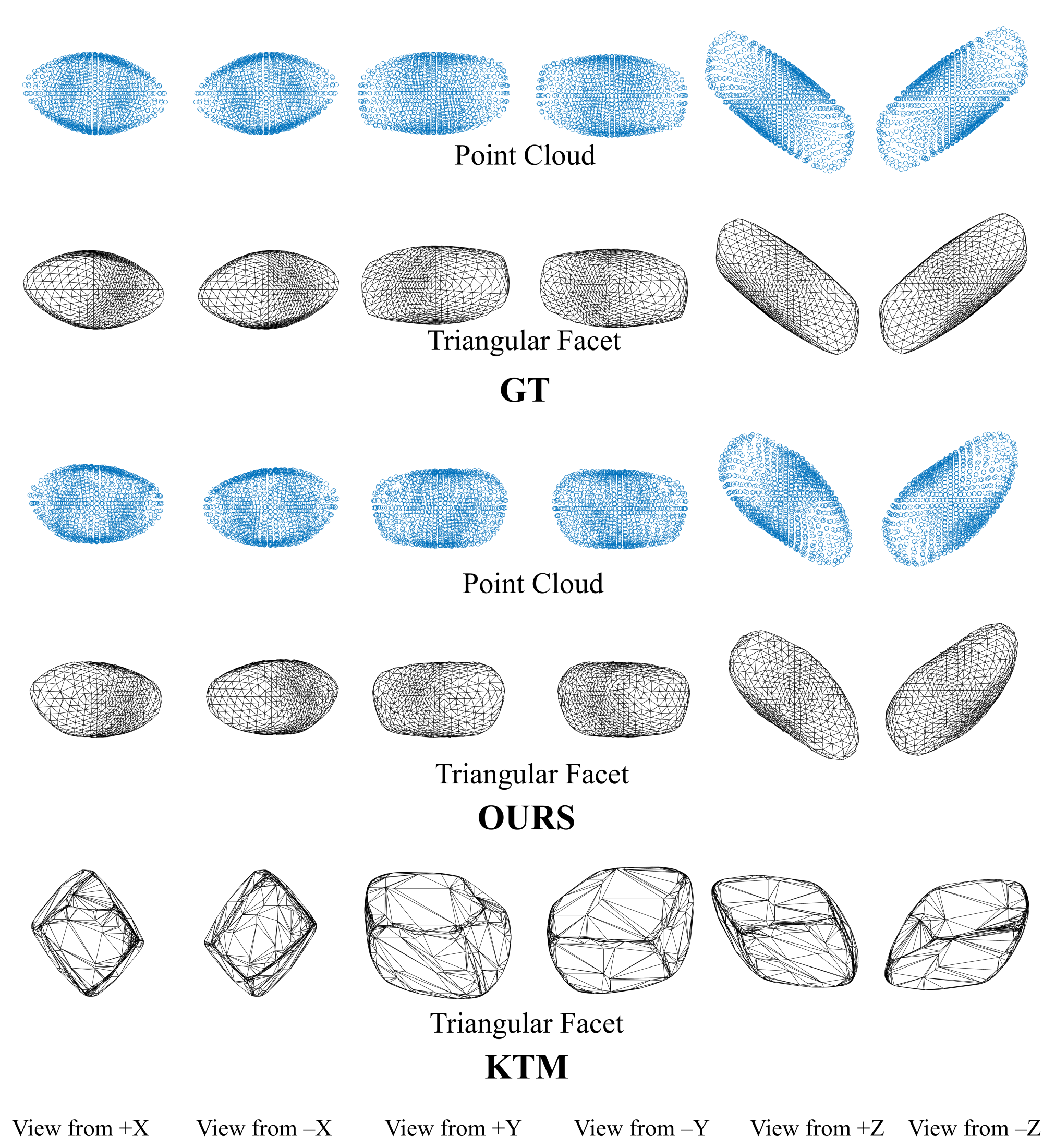}
        \captionsetup{justification=justified, singlelinecheck=false}
        \caption{Inversion results of rod-shaped model. The models are divided into three groups. The first group represents the ground truth of the 3D model, displayed using both point cloud and facet representations. The second group shows the 3D model reconstructed using our method, where the point cloud is converted into triangular facets for easier comparison with the KTM method. The third group represents the 3D model reconstructed using the KTM method. All models are aligned with the asteroid-centric coordinate system for consistency and clarity in comparison.  \label{fig:bang4}}
    \end{figure}

\begin{figure}[htbp]
        \centering
        \includegraphics[width=0.5\textwidth]{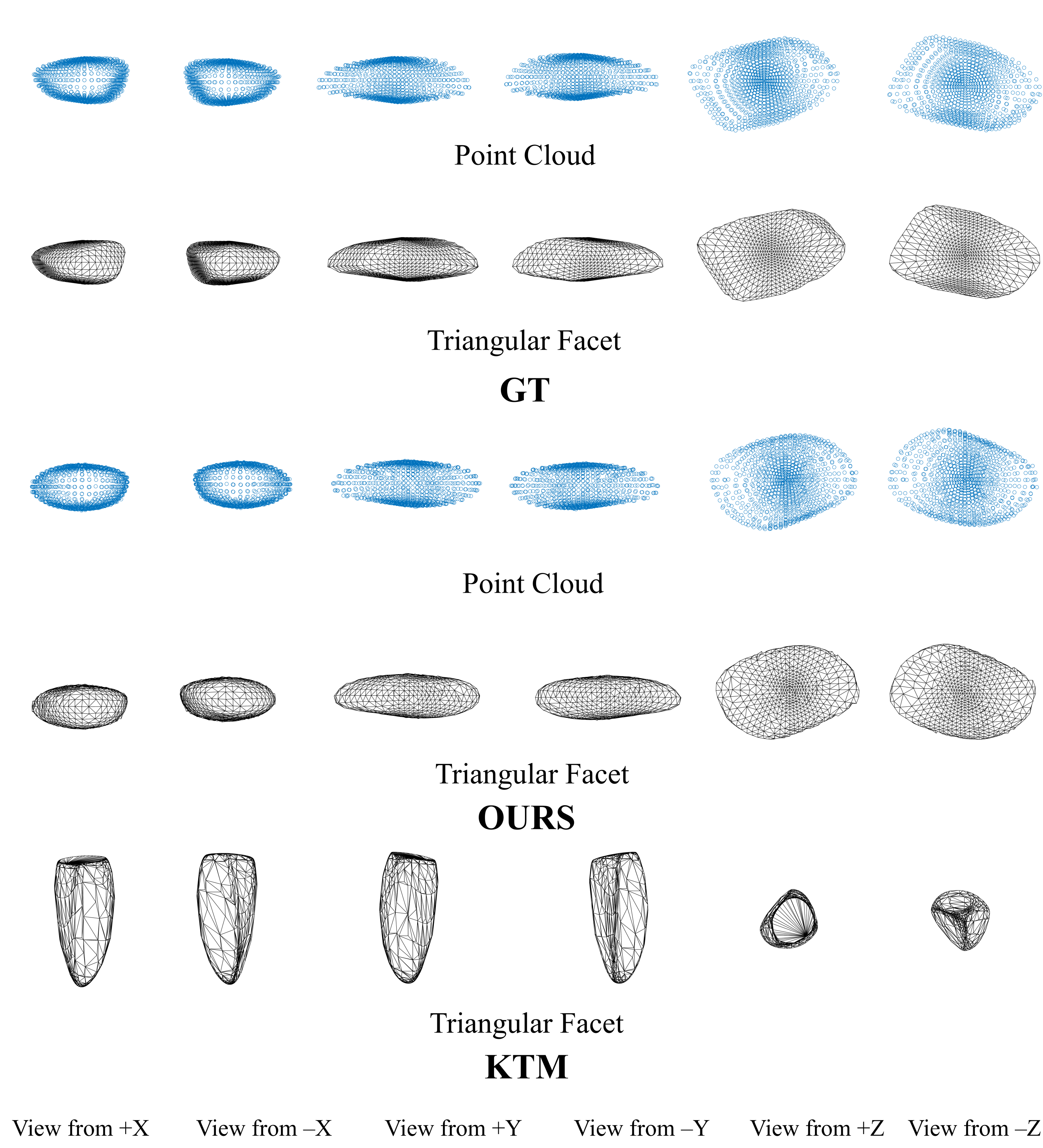}
        \captionsetup{justification=justified, singlelinecheck=false}
        \caption{Inversion results of sheet-like model. The models are divided into three groups. The first group represents the ground truth of the 3D model, displayed using both point cloud and facet representations. The second group shows the 3D model reconstructed using our method, where the point cloud is converted into triangular facets for easier comparison with the KTM method. The third group represents the 3D model reconstructed using the KTM method. All models are aligned with the asteroid-centric coordinate system for consistency and clarity in comparison.\label{fig:pian4}}

    \end{figure}

   To standardize the observation conditions, we used the actual observation conditions of asteroid 21689 (1999 RL38) as an example. In the KTM method, we fixed the pole axis and rotation period to the same values used in our approach, specifically $\lambda = 242^\circ$, $\beta = -55^\circ$, and a rotation period of 10.08457 hours. Fig.~\ref{fig:qiu4}-~\ref{fig:bang4} \& Fig.~\ref{fig:pian4} show the inversion results for spherical, rod-shaped, and disk-shaped 3D models, respectively. We note that all model images presented in this paper are aligned with the asteroid-centric coordinate system. We present the triangular mesh and 3D point cloud of the 3D models used for simulating the light curves, the triangular mesh and 3D point cloud predicted by the intelligent inversion method we proposed, and the results predicted by the \citet{2001Icar..153...24K}'s inversion algorithm, as shown in the figures.
        
    \begin{table}[!ht]
     \centering
     \captionsetup{justification=justified}
     \caption{Inversion performance comparison between KTM \citep{2001Icar..153...37K} and our method.}
     \begin{tabular*}{\columnwidth}{@{\extracolsep{\fill}}lcccc}
        \hline
        \multirow{2}{*}{model} & \multicolumn{2}{c}{CD/1e-3} & \multicolumn{2}{c}{time/s} \\
        & KTM & ours & KTM & ours  \\
        \hline
        spherical  & 3.605 & 1.250 & 45 & 0.56  \\
        rod-shaped & 176.53 & 4.004 & 50 & 0.56  \\
        sheet-like & 197.2 & 2.210 & 49 & 0.56  \\
        \hline
    \end{tabular*}
    \label{tab1}
    \end{table}

    Based on the above results, it is evident that the prediction model proposed in this study accurately captures the basic outline of the actual models. In contrast, the convex inversion algorithm \citep{2001Icar..153...37K} shows discrepancies, particularly with the disk-shaped 3D model, which may be due to the algorithm falling into local minima during optimization.
    
    More importantly, our method uses regression to make its prediction, which greatly improves its computational efficiency. It only takes our method 0.56s to predict a three-dimensional point cloud with 1024 vertices, as shown in the Table\ref{tab1}.

    During the experiment, we found that the prediction errors of the same 3D model varied significantly under different observation conditions. This variation is not only related to the number of sampling points but is also influenced by factors such as the distribution of sampling points. To investigate under what observation conditions of light curves our network can achieve better results, we defined a metric to quantify the sparsity of the light curve samples. It should be noted that this sparsity metric is specifically designed based on the size of the surface area covered by the observations. This metric takes into account the situation where certain regions have densely sampled points, leading to overlapping observations, and it also incorporates surface area weights to account for the different surface areas observed during a full rotation of the asteroid at different latitudes.
    
        \begin{figure}[!ht]
        \centering
        \includegraphics[width=0.5\textwidth]{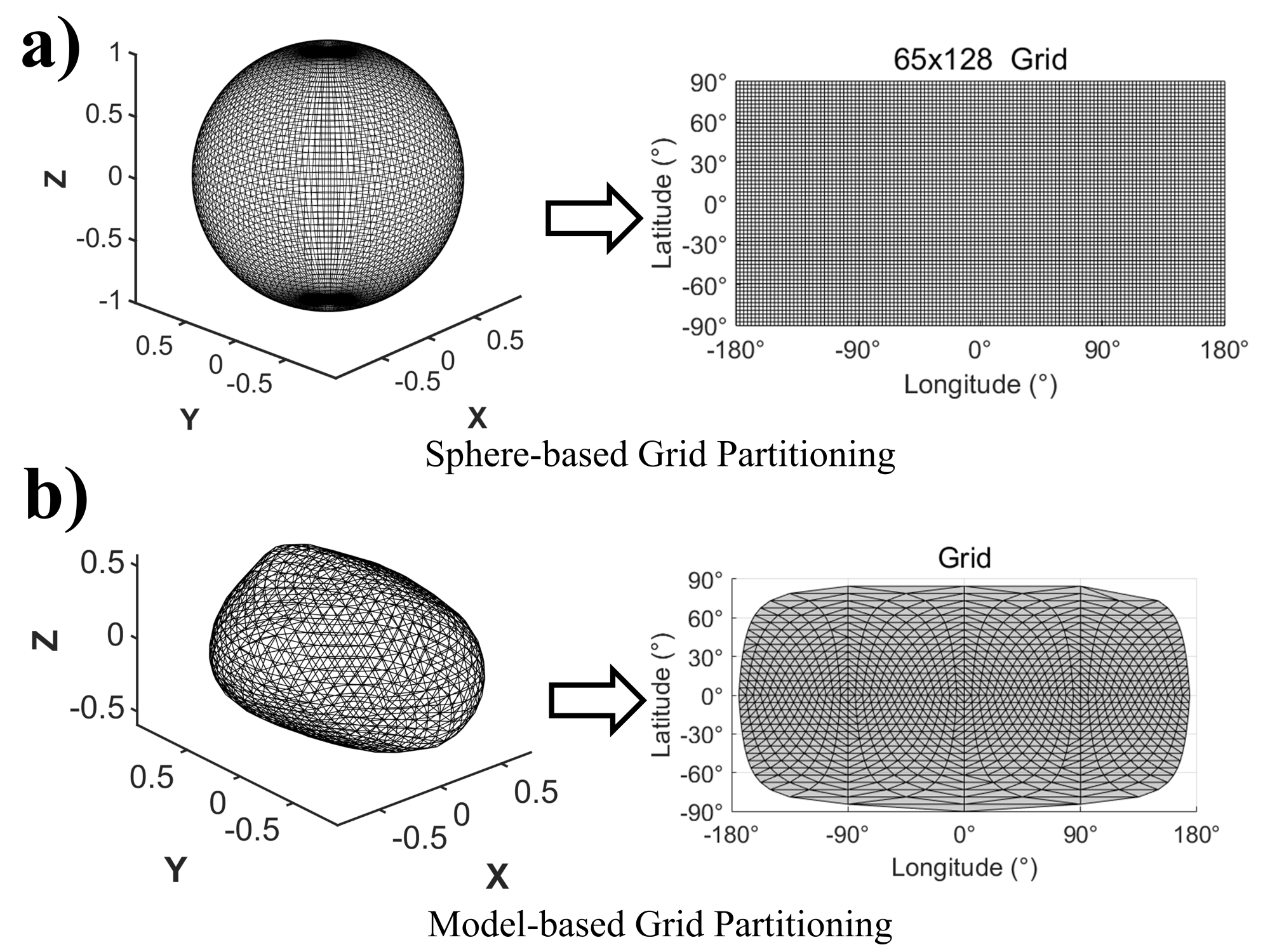}
        \captionsetup{justification=justified, singlelinecheck=false}
        \caption{Generation of the grids. Here a) represents a sphere-based grid partitioning method, and b) represents a model-based grid partitioning method.}
        \label{fig:f7}
    \end{figure}

   We divided the unit sphere's latitude and longitude into equal intervals, partitioning it into a 128*65 grid, as shown in part $a$ of Fig.\ref{fig:f7}. For each latitude direction $j$ (a total of 65 directions), we calculated the ideal scattering area of each grid in that latitude direction, resulting in the vector $x_{ideal}^{j}$: 
    \begin{equation}
   x_{ideal}^{j} =[x_{ideal,1,1}^{j},x_{ideal,1,2}^{j},x_{ideal,1,3}^{j},...,x_{ideal,k,l}^{j}],
   \end{equation}  
   
     Where $x_{ideal,k,l}^{j}$ represents the scattering area of each grid:
       \begin{equation}
       x_{ideal,k,l}^{j} = \max [Area_{k,l}\left (\frac{\mu_{j,i}}{\mu_{j,i} + 1} + c\mu_{j,i}  \right ) ].
       \end{equation}  
     
    We used the LS-L scattering law\citep{2001Icar..153...37K} to calculate the scattering area of each grid. The terms $k$ and $l$ represent the first $k$ grid from the top and the $l$ grid counterclockwise, $k\in \left\lbrack 1,65 \right\rbrack$,  $l\in \left\lbrack 1,128 \right\rbrack$. The term $\mu_{j,i}$ represents the cosine value of the angle between the grid normal vector in the $j-th$ latitude direction (from north to south) and the grid normal vector in the $i-th$ longitude direction (from west to east) and the surface normal vectors of the $k-th$ and $l-th$ grid，$i\in \left\lbrack 1,128 \right\rbrack$. When calculating the ideal scattering area in each latitude direction, we did not consider the influence of solar illumination, and we assumed that the cosine of the angle between the solar direction and the grid normal vector is always one. For each light curve sample, we first determined which of the 65 latitude divisions each sample point belongs to and then calculated the actual scattering area of each grid in each latitude direction $j$, resulting in the vector $x_{real}^{j}$ :
    \begin{equation}
  x_{real}^{j} =[x_{real,1,1}^{j},x_{real,1,2}^{j},x_{real,1,3}^{j},...,x_{real,k,l}^{j}].
   \end{equation} 
    
   Subsequently, we obtained the sparsity vector $x_{relative}$  for each latitude direction $j$ :
   \begin{equation}
    x_{relative}^{j} =[\frac{x_{real,1,1}^{j}}{x_{ideal,1,1}^{j}} ,\frac{x_{real,1,2}^{j}}{x_{ideal,1,2}^{j}},\frac{x_{real,1,3}^{j}}{x_{ideal,1,3}^{j}},...,\frac{x_{real,k,l}^{j}}{x_{ideal,k,l}^{j}}].
   \end{equation}
    The sparsity for each latitude direction $j$ can be defined as
    
    \begin{equation}
  \mathit{Sparsity_{j}(x)}  = l_{2}(x_{relative}^{j}) \in \left\lbrack 0,1 \right\rbrack,
    \end{equation}

\begin{figure}[!ht]
    \centering
    \includegraphics[width=0.5\textwidth]{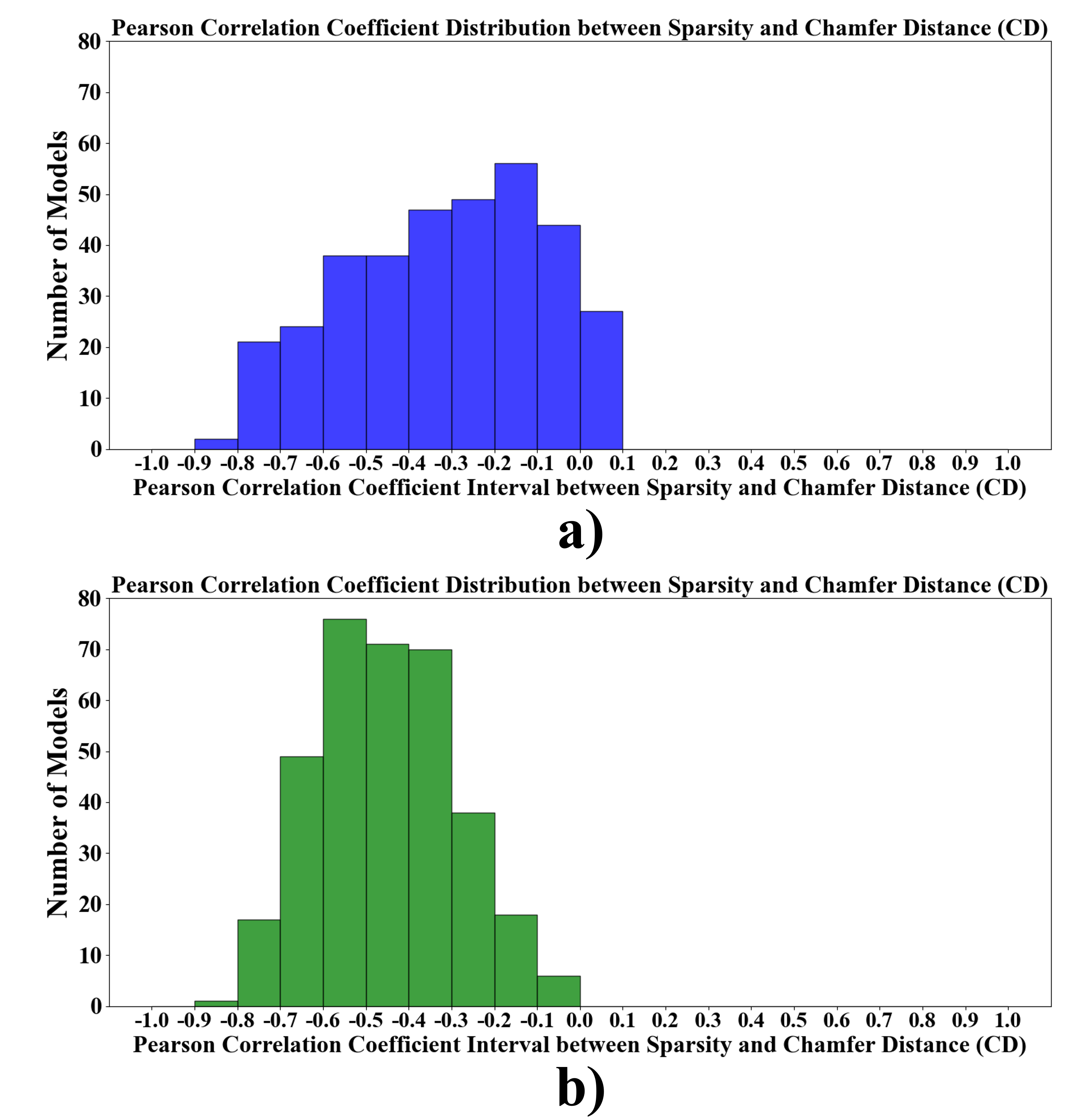}
    \captionsetup{justification=justified, singlelinecheck=false}
    \caption{Distribution of the Pearson correlation coefficient between sparsity and CD. Panel a)  shows the distribution of the Pearson correlation coefficient between sparsity and CD calculated using the sphere-based grid partitioning method for sparsity. Panel b) shows the distribution of the Pearson correlation coefficient between sparsity and CD calculated using the model-based grid partitioning method for sparsity.}
    \label{fig:f8}
\end{figure}
   where the L-p norm of the vector x is given by

    \begin{equation}
    l_{p}(x) = \left( {\sum_{i = 1}^{n}\left| x_{i} \right|^{p}} \right)^{1/p}.
    \end{equation}
    
   Here, $p$ is a parameter used to calculate the norm of a vector. We calculated the total surface area $Area_j$ of the grids in each latitude direction $j$, normalized this value, and used the normalized result as the area weight $W_j$:
     \begin{equation}
    W_j = \frac{Area_j}{4\pi}.
    \end{equation}

    Finally, our Sparsity is defined as
    \begin{equation}
   \mathit{Sparsity} = \sum_{j=1}^{65}{\mathit{{Sparsity}_j}}\times{W_j}.
    \label{eq:14}
    \end{equation}
    
   We calculated the sparsity of the light curve samples for 346 models in the test set using Equation~\eqref{eq:14}. Then, using Equation~\eqref{eq:10}, we calculated the CD between the inversion results and the ground truth as a measure of inversion accuracy error. For each model, we computed the Pearson correlation coefficient\citep{cohen2009pearson} between the sparsity of its light curve samples and the CD between the inversion results and the ground truth, and plotted the distribution of the Pearson correlation coefficients, as shown in part $a)$ of Fig.\ref{fig:f8}.

 In Fig.\ref{fig:f8}, part $a)$ shows that the Pearson correlation coefficients are generally less than zero, indicating a negative correlation between sparsity and CD error. The distribution of Pearson correlation coefficients is relatively balanced. For some models, there is a strong correlation between sparsity and CD error, while for others, the correlation is weak. We consider that the sparsity definition using the sphere-based grid partitioning method may not be fair for certain models. Therefore, we introduced the model-based grid partitioning method, where the triangular facets of the 3D model are used to create the grid, as shown in part $b)$ of Fig.\ref{fig:f7}. The sparsity is then calculated using the same method. Using this revised approach, we computed the sparsity of the light curve samples for 346 models from the test set and plotted the distribution of the Pearson correlation coefficients between sparsity and CD error, as shown in part $b)$ of Fig.\ref{fig:f8}.

 The results shown in Fig.\ref{fig:f8} indicate that the sparsity calculated using the model-based grid partitioning method exhibits a stronger correlation with CD. In general, the greater the sparsity of the light curve samples, the higher the accuracy of the inverted model. Therefore, sparsity can be used as an indicator to assess the quality of light curve sample data and to optimize the trajectory design of near-Earth spacecraft and light curve sampling strategies. This indicator ensures that the collected data more effectively supports the light curve inversion process, thereby improving the accuracy of the generated 3D models of asteroids.

 \subsection{Simulation of the process of the probe approaching an asteroid}
   During space missions exploring asteroids, as a probe approaches its target asteroid, the quality and quantity of the data collected increases. Theoretically, this can improve the inversion accuracy of the asteroid 3D model. Different from traditional inversion methods, which require numerical re-optimization when encountering new data and can be time-consuming, our proposed method leverages its speed advantage to quickly utilize new data and provide a model that is more accurate than in the previous stage, enabling rapid on-board autonomous decision-making.
  
\begin{figure}[htbp]
            \centering
            \includegraphics[width=0.5\textwidth]{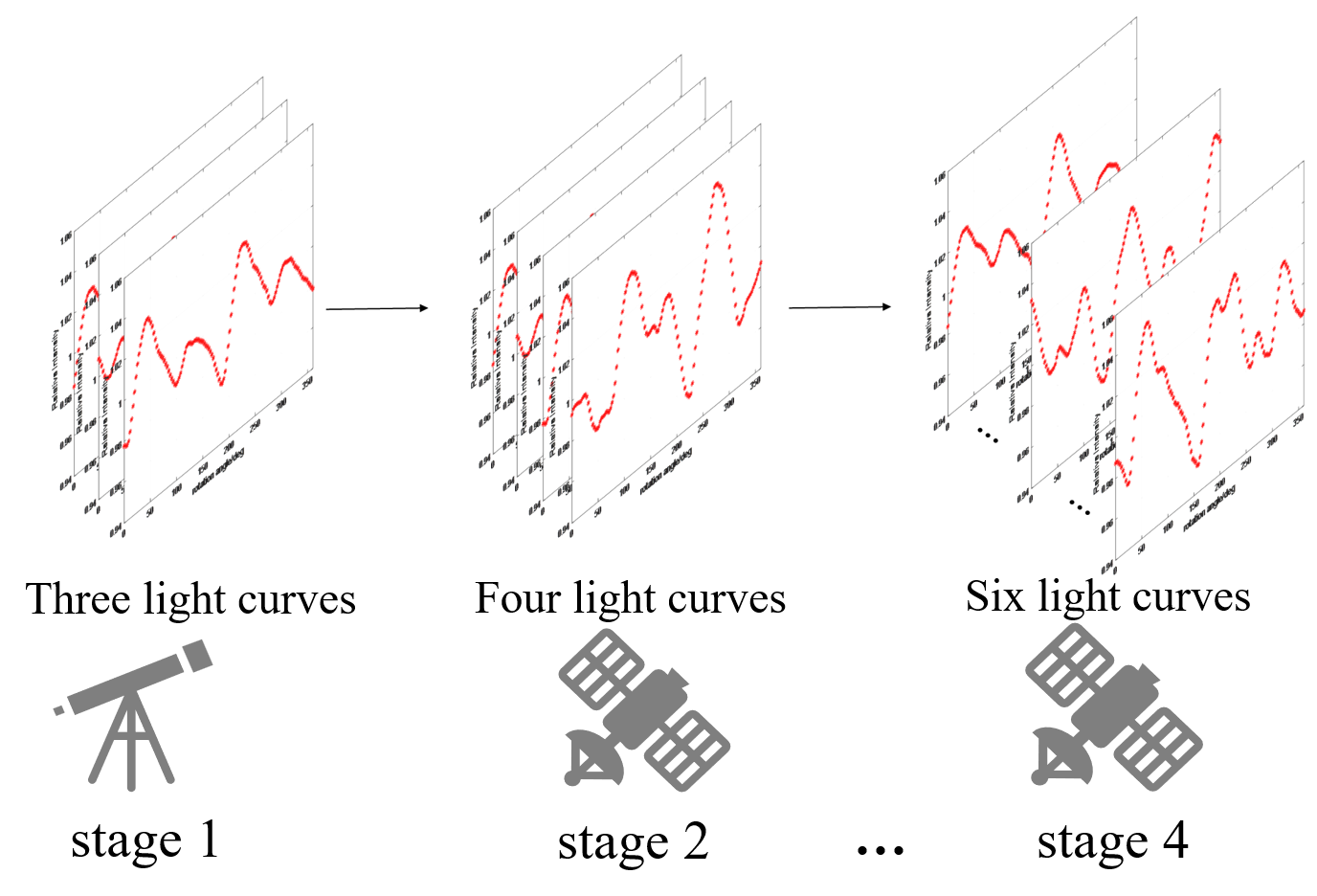}
            \captionsetup{justification=justified, singlelinecheck=false}
            \caption{Schematic of the simulated approach detection. Using telescopes on the ground and optical payloads on the spacecraft after launch, the light curve data increases gradually during the approach process, starting with 
            three light curves in Stage 1 and adding one additional light curve at each subsequent stage.}
            \label{fig:f99}
       \end{figure}

\begin{figure}[!ht]
    \centering
    \begin{subfigure}[h]{0.5\textwidth}
        \centering
        \includegraphics[width=\textwidth]{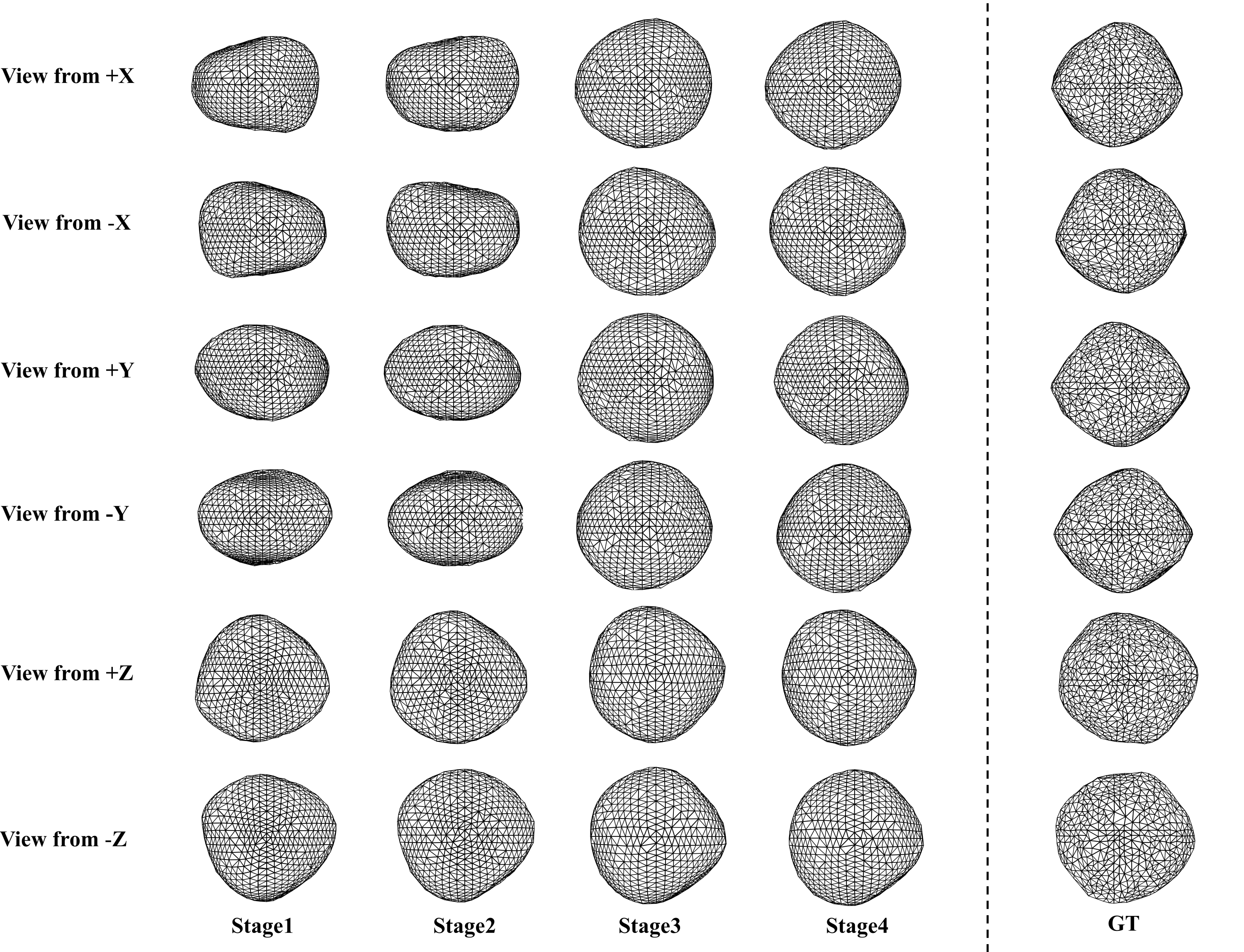}
        \captionsetup{justification=justified, singlelinecheck=false}
        \subcaption{ }

        \label{fig:f9a}
    \end{subfigure}
    \hfill
    \begin{subfigure}[h]{0.5\textwidth}
        \centering
        \includegraphics[width=\textwidth]{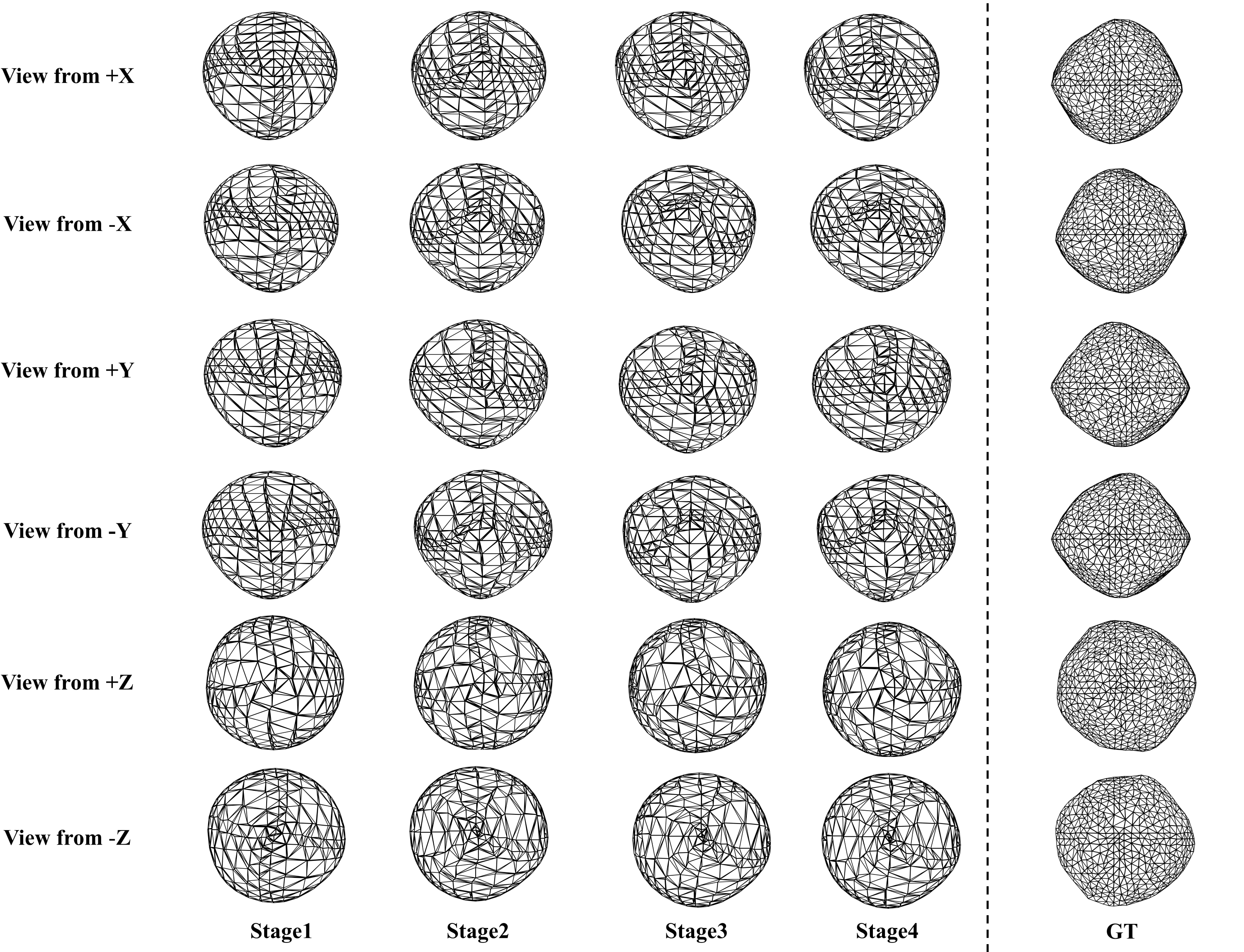}
        \captionsetup{justification=justified, singlelinecheck=false}
        \subcaption{ }
        
        \label{fig:f9b}
    \end{subfigure}
    \captionsetup{justification=justified, singlelinecheck=false}
    \caption{Inversion results of simulated approach detection process for asteroid 101955 Bennu. Panel (a): Results using our method. Each row of images is labeled with the corresponding observation perspective. The first column shows the inversion results based on the ground observation light curve, while the second to fourth columns display the inversion results after progressively adding light curve data collected by the detector during the flight phase. The fifth column presents the actual 3D model of 101955 Bennu. Panel (b): Corresponding results using the KTM method.}
    \label{fig:f9}
\end{figure}

   We used the flight trajectory data of OSIRIS-REx\citep{nasa} to simulate the spacecraft's approach to the asteroid 101955 Bennu. First, we simulated three ground-based observations every two months. Each ground-based observation was designed to simulate the spacecraft observing the asteroid for the duration of one complete rotation. Then, every two months after the launch of the probe, we simulated the start-up of the probe and observed the luminosity of the asteroid rotation for one period, a total of three times, as shown in the schematic diagram in Fig.\ref{fig:f99}. It can be seen from the inversion results in Fig.\ref{fig:f9a} that the inversion results of the first column, based on the ground observation data, have a large deviation from the actual situation. With the launch of the detector, when two light curves are added, the inversion results(shown in the third column) more accurately reflect the basic contour of the real model. In general, as the amount of data increases during the simulation process, the inversion accuracy improves.

 Additionally, we compared our method with the KTM method. In the case of the KTM method, we fixed the pole axis and rotation period to the same values used in our model, specifically $\lambda = 45^\circ$, $\beta = -88^\circ$, and a rotation period of 4.2975 hours. The inversion results using the KTM method are shown in Fig.\ref{fig:f9b}. We compared the inversion results of different methods at various stages, using the CD between the inverted models and the ground truth, as shown in Table.\ref{cdbennu}. It can be observed that in Stage 2-4, although the number of light curve points increases, the CD between the models inverted by the KTM method and the ground truth no longer decreases, stabilizing around 137 (1e-3). In contrast, our method shows improvement, with the CD of the inverted models decreasing from 152 (1e-3) to 127 (1e-3) in Stage 3-4 and the accuracy of the model inverted in Stage 4 surpassing that of the KTM method. Our method demonstrates a clear trend of reduced error, and we expect that there is still potential for further improvement in accuracy as the data volume increases further.

\begin{table}[!ht]
    \centering
    \captionsetup{justification=justified}
    \caption{CD between the inverted model and the ground truth at different stages of the approach process.}
     \begin{tabular*}{\columnwidth}{@{\extracolsep{\fill}}lcccc}
        \hline
        \multirow{2}{*}{method} & \multicolumn{4}{c}{CD/1e-3} \\ 
        & Stage1 & Stage2 & Stage3 & Stage4 \\
        \hline
        KTM  & 144 & 137 & 139 & 137 \\
        ours & 163 & 151 & 152 & 127\\       
        \hline
     \end{tabular*}
     \label{cdbennu}  
\end{table}

    \subsection{ Real light curve experiments}
    The observational data were selected from the historical observation data of asteroids 3337 Miloš and 1289 Kutaïssi available on the DAMIT website\citep{2010A&A...513A..46D}. These data were obtained from the Lowell Observatory survey\citep{2014MPBu...41..286K} conducted between 1990 and 2012. The phase angle range for the observational data of asteroid 3337 Miloš is from 0.19$^\circ$ to 22.59$^\circ$, and for asteroid 1289 Kutaïssi, it is from 0.31$^\circ$ to 21.61$^\circ$. These two asteroids have extensive sampling points, and their historical photometric data have already been normalized, resulting in high-quality data. Therefore, the proposed deep learning network inversion model was used to predict the 3D shapes of these two asteroids, and the results were compared with those obtained using the \citet{2001Icar..153...24K} inversion program code provided by the DAMIT website.

\begin{figure}[!ht]
    \centering
    \begin{subfigure}[h]{0.51\textwidth}
        \centering
        \includegraphics[width=\textwidth]{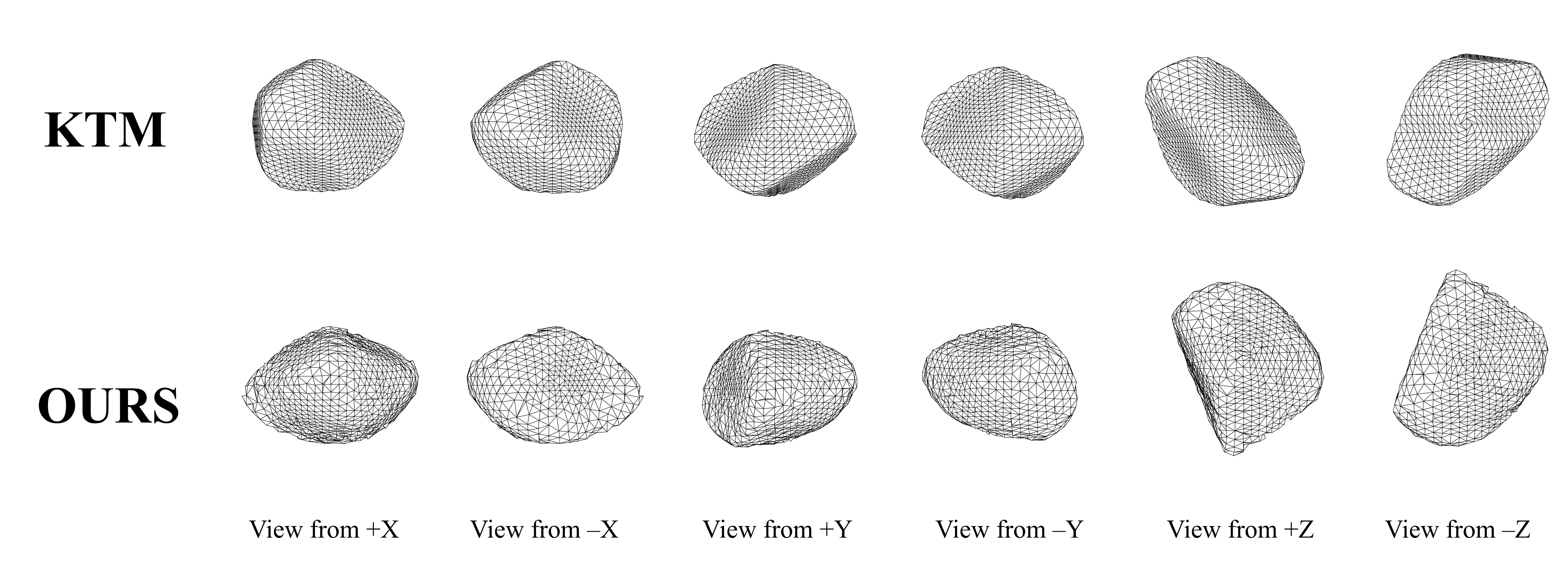}
        \captionsetup{justification=justified, singlelinecheck=false}
        \subcaption{ }

        \label{fig:f11a}
    \end{subfigure}
    \hfill
    \begin{subfigure}[h]{0.51\textwidth}
        \centering
        \includegraphics[width=\textwidth]{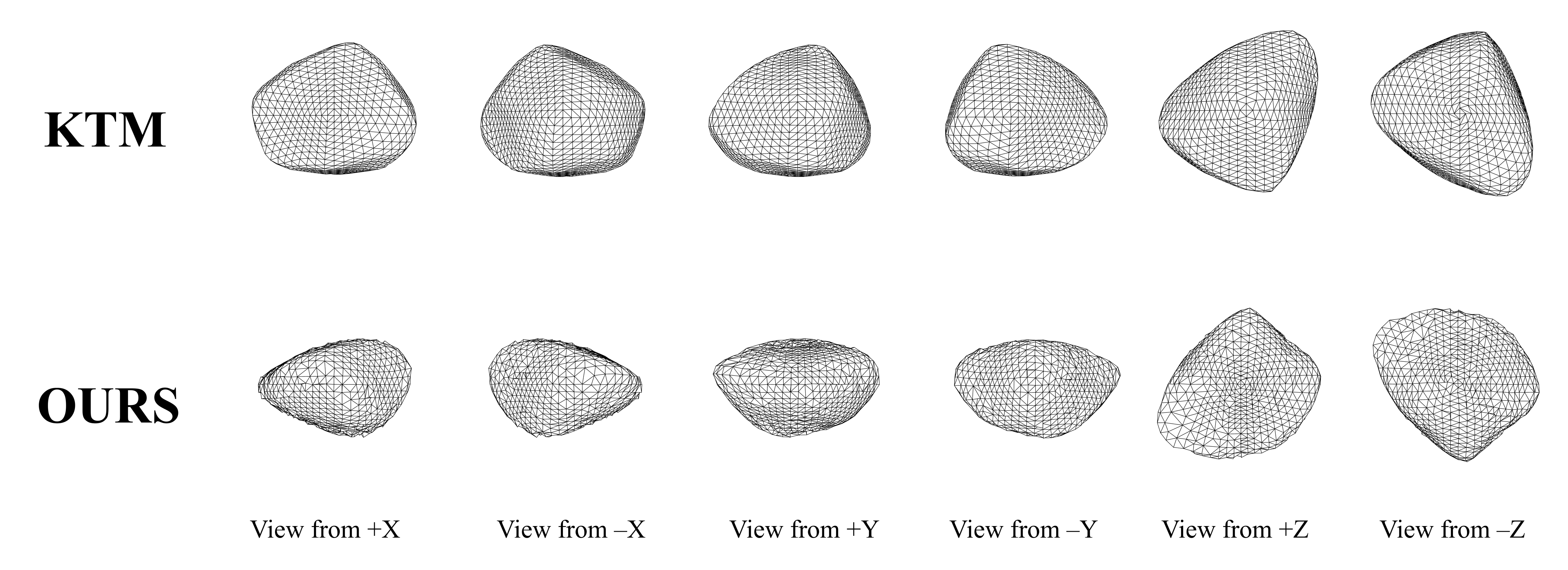}
        \captionsetup{justification=justified, singlelinecheck=false}
        \subcaption{ }
        \captionsetup{justification=justified, singlelinecheck=false}
        \label{fig:f11b}
    \end{subfigure}
    \captionsetup{justification=justified, singlelinecheck=false}
    \caption{Convex inversion results using real light curve data. Panel (a): Results for asteroid 3337 Miloš. The top row shows the method of \citep{2001Icar..153...37K}, and bottom row shows our method. Panel (b): Results for asteroid 1289 Kutaïssi. The real shapes are unknown due to a lack of close observations.}
    \label{fig:f11}
\end{figure}

    Fig.\ref{fig:f11a} and Fig.\ref{fig:f11b} show the inversion results for asteroids 3337 Miloš and 1289 Kutaïssi, respectively. From the figures, it can be observed that the prediction results of the two methods differ significantly. Due to a lack of ground truth, it is hard to determine which  method gives better results. We conducted light curve fitting experiments on the 3D models inverted from asteroids 3337 Miloš and 1289 Kutaïssi using our method and compared them with the 3D models inverted using the \citet{2001Icar..153...24K}'s  method. Since the observed light curve data for these two asteroids spans a long period, we used the phase angle as the horizontal axis in order to plot the graphs and examine the fitting results in the light curve fitting experiments. The light curve fitting results for asteroids 3337 Miloš and 1289 Kutaïssi obtained by our method and the KTM method are shown in Fig.\ref{fig:milos} and Fig.\ref{fig:kutai}. It can be seen that the models produced by our method and the KTM method yield similar results in light curve fitting. We calculated the mean absolute error in light curve fitting for the 3D models obtained by different methods, as shown in Table\ref{taberror}.
    
 \begin{table}[htb]
     \centering
     \captionsetup{justification=justified}
     \caption{Mean absolute error in light curve fitting between KTM\citep{2001Icar..153...37K} and our method.}
     \begin{tabular*}{\columnwidth}{@{\extracolsep{\fill}}lcccc}
        \hline
        \multirow{2}{*}{Asteroid} & \multicolumn{2}{c}{Mean absolute error in light curve fitting}\\
        &Model from KTM & Model from ours   \\
        \hline
        3337 Miloš  & 0.164 & 0.173 \\
        1289 Kutaïssi   & 0.092 & 0.097  \\
        \hline
    \end{tabular*}
    \label{taberror}
    \end{table}

     \subsection{Convex hull inversion of non-convex model}
\begin{figure}[htbp] 
    \centering
    \begin{subfigure}[h]{0.5\textwidth}
        \centering
        \includegraphics[width=\textwidth]{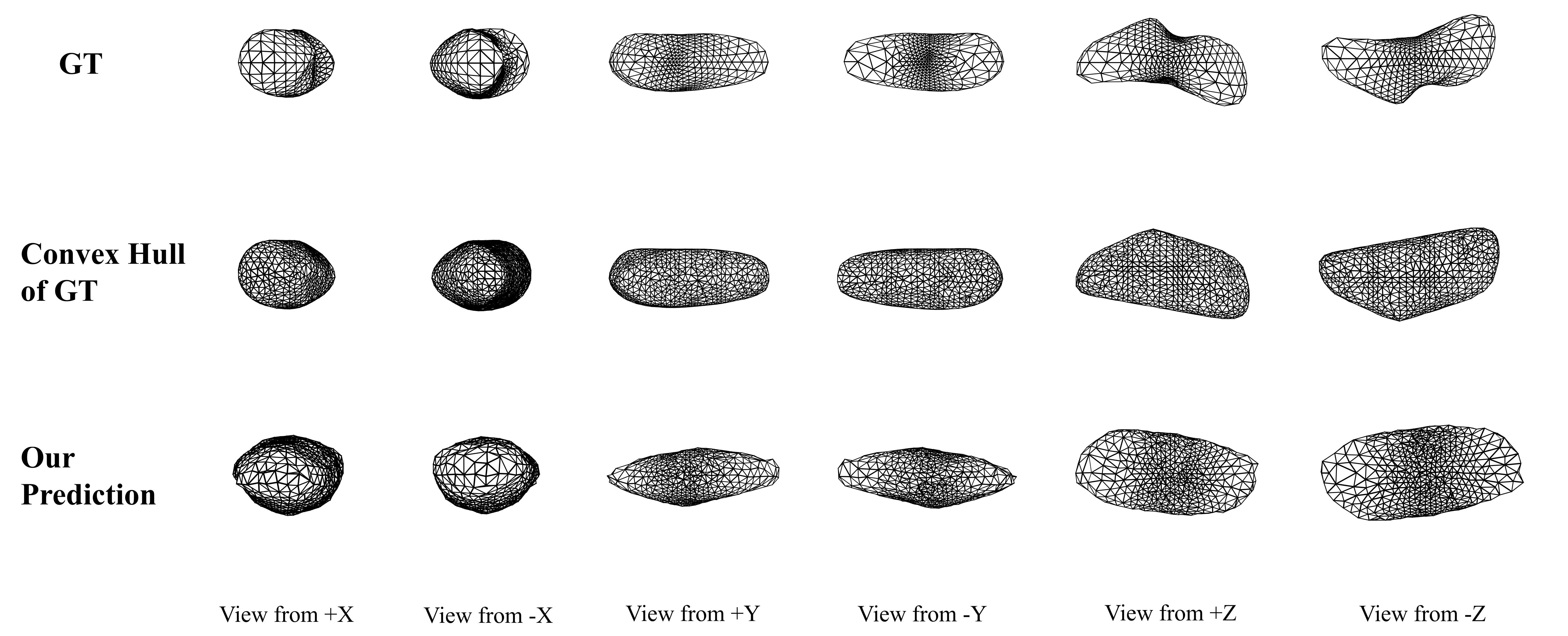}
        \captionsetup{justification=justified, singlelinecheck=false}
        \subcaption{ }
        \label{fig:f10a}
    \end{subfigure}
    \hfill
    \begin{subfigure}[h]{0.5\textwidth}
        \centering
        \includegraphics[width=\textwidth]{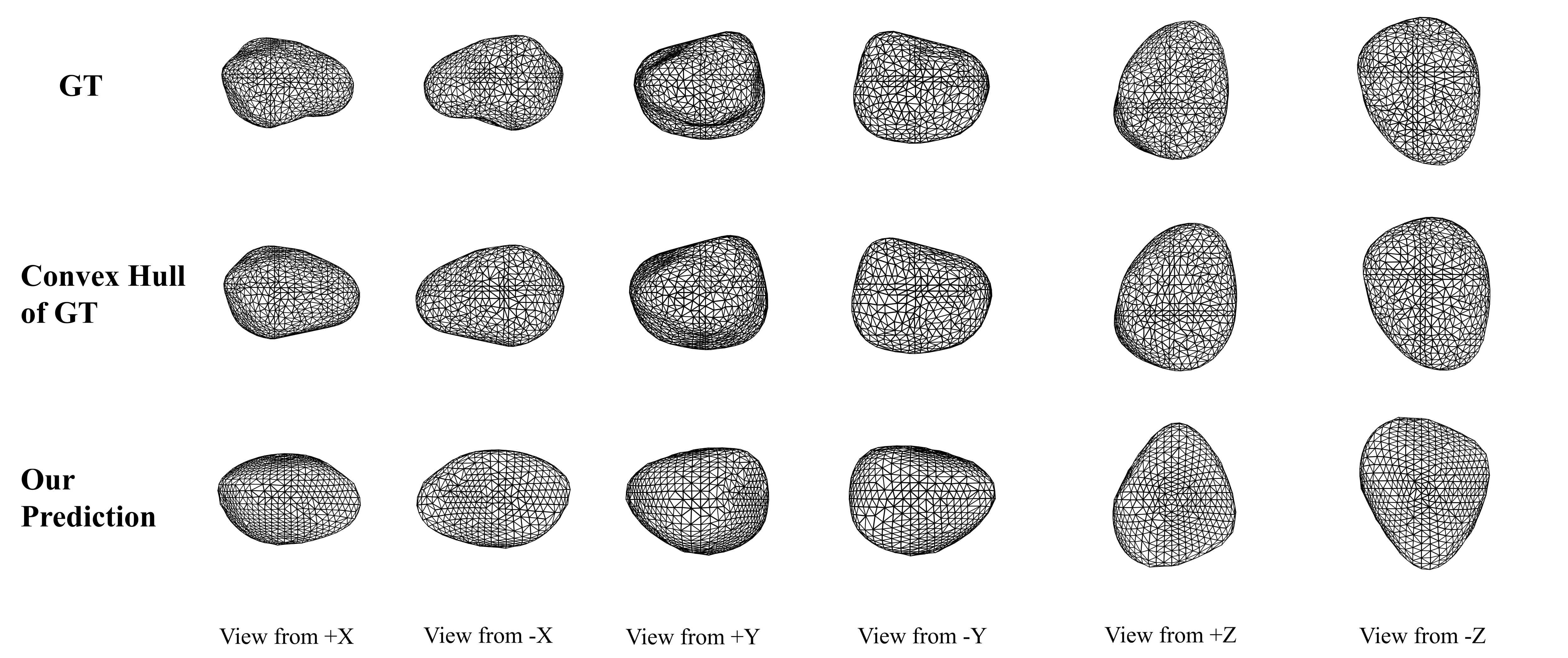}
        \captionsetup{justification=justified, singlelinecheck=false}
        \subcaption{ }
        \label{fig:f10b}
    \end{subfigure}
    \captionsetup{justification=justified, singlelinecheck=false}
    \caption{Convex hull inversion results for non-convex asteroids. Panel (a): Asteroid 433 Eros. The top row shows the NEAR Shoemaker mission shape model \citep{zuber2000shape}, the middle row shows its convex hull, and the bottom row shows our inversion result. Panel (b): Asteroid 113 Lutetia. The top row shows the shape model by \citep{farnham2013shape}, the middle rows shows its convex hull, and the bottom row shows our inversion result.}
    \label{fig:f10}
\end{figure}

    The 3D models we used when training the network are convex, which is inconsistent with the actual three-dimensional shape of the asteroids. To verify the applicability of our proposed method to non-convex models, we simulated the light curves of non-convex 3D models based on the 433 Eros and 113 Lutetia asteroids, as shown in Fig.\ref{fig:f10a} and Fig.\ref{fig:f10b}. The results from the figures show that although we used the light curve data of the non-convex models, we are still able to invert a model that closely approximates the shape of the minimum circumscribed convex body of the non-convex model. 
    
   We conducted a light curve fitting experiment on our inversion results for 433 Eros and 113 Lutetia, and we computed the fitting errors. The average absolute error for the light curve fitting of 433 Eros is 0.0641, while for 113 Lutetia, the average absolute error is 0.0179. Some of the experimental results are presented in Fig.\ref{fig:f15}. It can be observed that for the small-area non-convex 3D model of 113 Lutetia, the light curve fitting results are quite good. However, for the large-area non-convex 3D model of 433 Eros, the fitting is slightly less accurate, though the shape of the inverted model still closely matches its minimum circumscribed convex body. These results can provide effective initial values for some numerical optimization algorithms, such as SAGE, thereby reducing inversion time.
   
  For the current  concave areas determination methods, the inversion accuracy of convex hulls that we obtained remains unacceptable. It still leads to the photometric differences that represent the shadowing effects of non-convex areas being affected by the shape differences between the inverted convex hull and the true convex hull. As a result, we have not yet conducted non-convex region detection experiments on them.

\subsection{Determination of concave areas results}
\begin{figure}[htbp]
            \centering
            \includegraphics[width=0.5\textwidth]{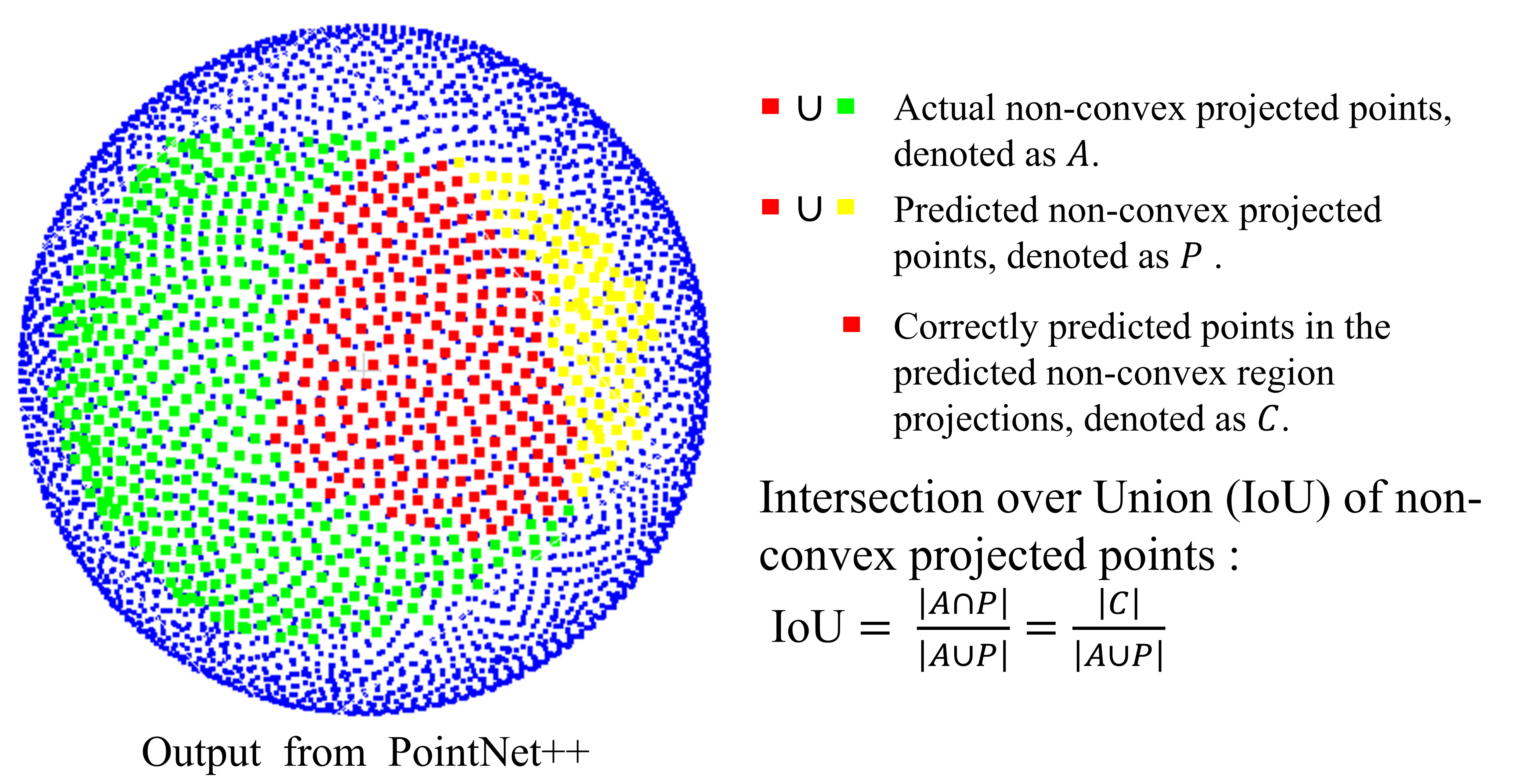}
            \captionsetup{justification=justified, singlelinecheck=false}
            \caption{Illustration of the Intersection over Union (IoU)\citep{qi2017pointnet} calculation in the non-convex region prediction task.}
            \label{fig:IOU}
       \end{figure}

\begin{figure}[htb]
        \centering
        \includegraphics[width=0.5\textwidth]{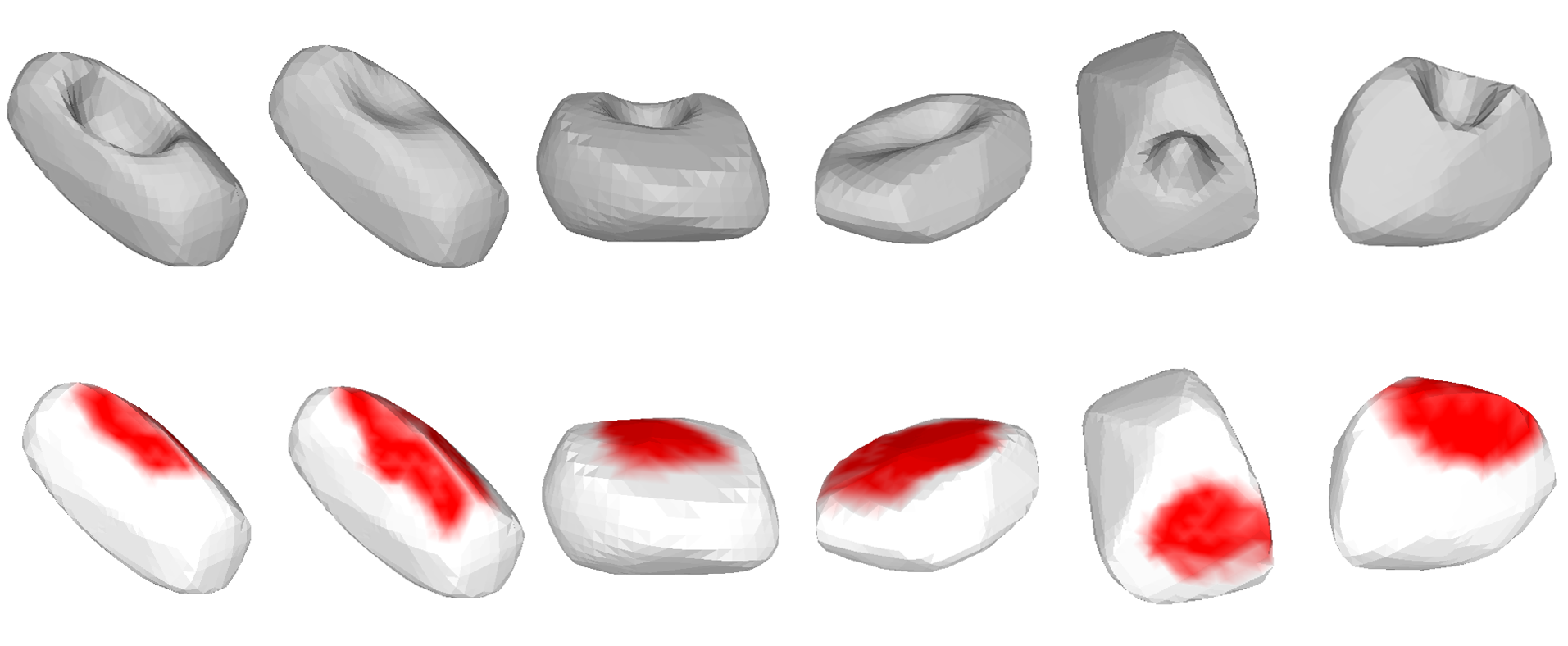}
        \captionsetup{justification=justified, singlelinecheck=false}
        \caption{Prediction results of non-convex areas under real observation conditions. The top shows the true non-convex model, and the bottom shows the prediction results on the  convex hull, with the red color indicating the predicted concave area.}
        \label{fig:n3}
    \end{figure}

    Using the simulated light curve of the non-convex model under real observation conditions, we predicted the region of depression on the minimum circumscribed convex hull of the non-convex model. For the concave areas discrimination network,  we used the intersection over union (IoU) metric\citep{qi2017pointnet} to evaluate the prediction results. A diagram illustrating the IoU is shown in Fig.\ref{fig:IOU}. If the non-convex surface in the model is large, the IoU of the predicted non convex region point cloud can reach 0.89. For the smaller concave areas, the predicted region will be slightly offset, and as shown in  Fig.\ref{fig:n3}, its intersection union ratio is about 0.6.

\section{\textbf{Conclusions}}
         We have proposed an asteroid 3D shape inversion algorithm based on convolutional and transformer networks that establishes and learns the mapping relationship between asteroid photometry and geometric position coordinates in relation to the 3D point cloud distribution of the asteroid by using deep neural networks. On a RTXA5000 GPU, it can predict the 3D point cloud of the convex hull of an asteroid in just 0.56s, compared to the 45-second runtime of traditional optimization algorithms, thus significantly enhancing inversion efficiency.

         For non-convex asteroids, we proposed a method based on the convex hull to predict concave areas. Although this method has only been implemented in simulations so far, it still reveals the potential of deep learning technology in the inversion of non-convex three-dimensional shapes of asteroids. In the next step, we will attempt to apply the concave areas prediction algorithm to real observational data. Whether using the \citet{2001Icar..153...24K} method or our transformer-based approach to predict the shape of non-convex asteroids, there will be differences compared to their strict convex hull, which will manifest in the photometric data. This is particularly challenging, especially for asteroids with a small size of nonconvexity\citep{2003A&A...404..709D}. Once the concave areas are identified, the size and depth of the craters will be aspects that require our attention. If we can roughly determine the shape of a non-convex asteroid, the prediction of other physical properties, including mass, center of mass, and rotation axis, will become more accurate.

\section*{\textbf{Acknowledgements}}
This research was funded by the National Science and Technology Major Project (2022ZD0117401) and the National Defense Science and Technology Innovation Special Zone Project Foundation of China grant number 19-163-21-TS-001-067-01.


\clearpage
\setcounter{figure}{0}
\renewcommand{\thefigure}{A\arabic{figure}}
\section*{Appendix A: Additional figures}

\begin{figure*}[!hb]
    \centering
    \includegraphics[width=\linewidth]{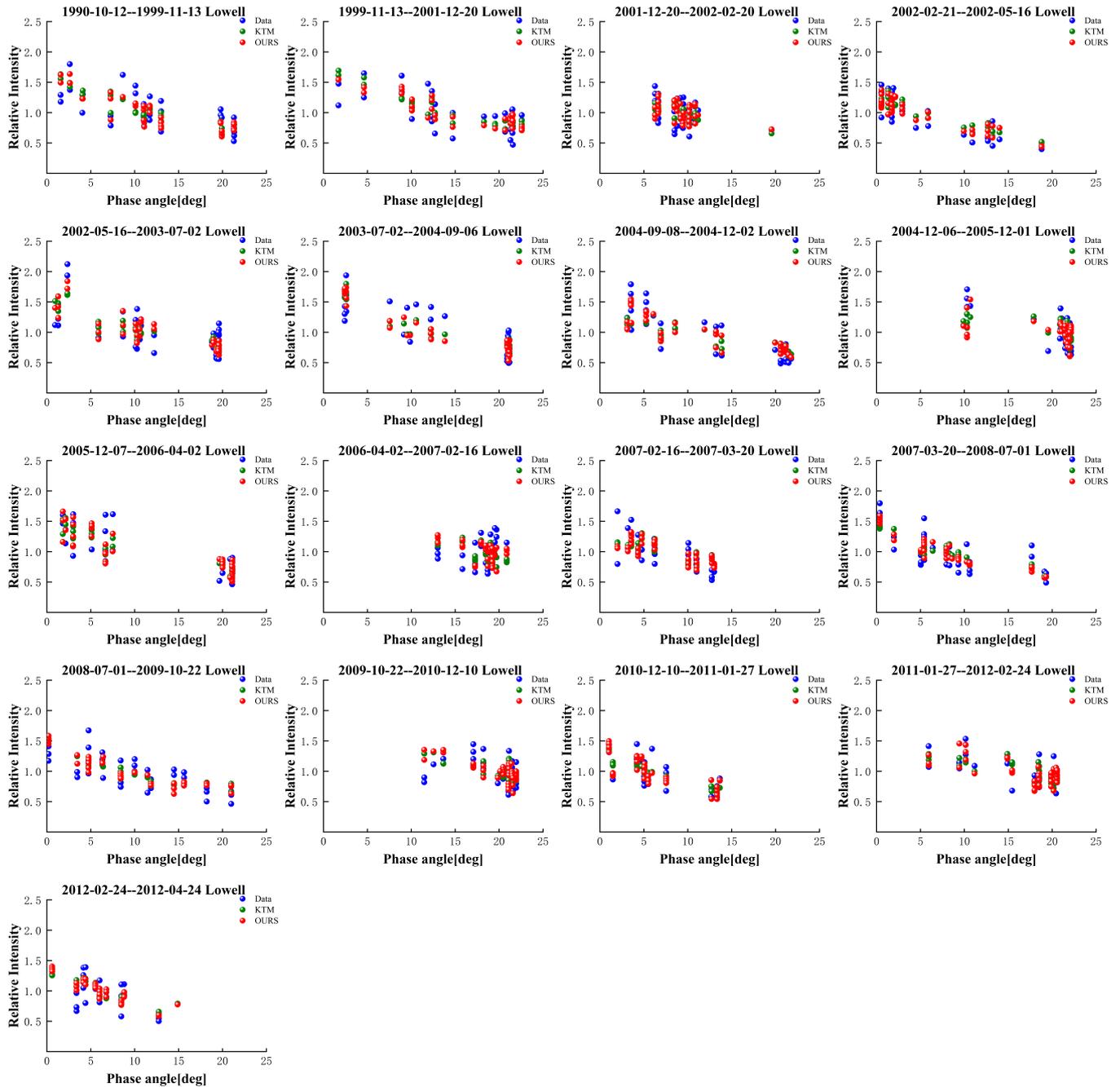}
    \captionsetup{justification=justified, singlelinecheck=false}
    \caption{Light curve fitting results of the 3D model of asteroid 3337 Miloš obtained by KTM \citep{2001Icar..153...24K} and our method, which was collected from the Lowell Observatory between October 12, 1990, and April 24, 2012. We indicate the time period of the observational data above each chart. The blue points represent the observational data, the green points represent the simulated brightness from the inverted model using the KTM, and the red points represent the simulated brightness from the inverted model using our method. The horizontal axis represents the phase angle, while the vertical axis represents the relative brightness.}\label{fig:milos}
\end{figure*}

\begin{figure*}[!ht]
    \centering
    \includegraphics[width=\linewidth]{kutai-revise.png}
    \captionsetup{justification=justified, singlelinecheck=false}
    \caption{Light curve fitting results of the 3D model of asteroid  1289 Kutaïssi obtained by KTM \citep{2001Icar..153...24K} and our method, which was collected from the Lowell Observatory between November 10, 1998, and March 16, 2012. We indicate the time period of the observational data above each chart. The blue points represent the observational data, the green points represent the simulated brightness from the inverted model using the KTM, and the red points represent the simulated brightness from the inverted model using our method. The horizontal axis represents the phase angle, while the vertical axis represents the relative brightness.}
    \label{fig:kutai}
\end{figure*}

\clearpage   
 \begin{figure*}[!ht]
        \centering
    \includegraphics[width=\textwidth]{eros_fitting.png}
       \captionsetup{justification=justified, singlelinecheck=false}
       \caption*{(a)}   
        \label{fig:eros} 
    \end{figure*}

    \begin{figure*}[!ht]
    \centering
    \includegraphics[width=\textwidth]{lutetia_fitting.png}
       \captionsetup{justification=justified, singlelinecheck=false}
    \caption*{(b)}
        \label{fig:lutetia}
    \caption{Partial light curve fitting results of the 3D models of 433 Eros and 113 Lutetia obtained by our method. Panel (a) represents the results for 433 Eros, while Panel (b) represents the results for 113 Lutetia. The blue points represent the simulated observational data, while the red triangles represent the simulated brightness of the inverted models. The horizontal axis represents the phase of rotation, and the vertical axis represents the relative brightness.}
    \label{fig:f15}
\end{figure*}

\end{CJK}

\end{document}